\RequirePackage[T1]{fontenc}
\documentclass[12pt]{article}
\pdfoutput=1 
\usepackage[height=8.85in,width=6.45in]{geometry}

\usepackage[utf8]{inputenc}
\usepackage{amsmath}
\usepackage{amssymb}
\usepackage{mathtools}
\numberwithin{equation}{section}
\usepackage{slashed}
\usepackage{braket}
\usepackage[svgnames]{xcolor}
\usepackage[colorlinks,linktocpage=true,citecolor=DarkGreen,linkcolor=FireBrick]{hyperref}
\usepackage{cite}
\usepackage{graphicx}
\usepackage{tikz}
\usetikzlibrary{decorations.markings}
\usepackage{tikz-cd}
\usepackage{times}
\usepackage[scaled]{couriers}
\usepackage{bm}
\usepackage{subfig}
\usepackage{mathrsfs}
\usepackage{amsthm}

\usepackage{tikz}
\usetikzlibrary{arrows.meta}

\usepackage{xcolor}
\usepackage{mdframed}

\renewenvironment{figure}[1][]{
  \begin{originalfigure}[#1]
    \begin{mdframed}[linecolor=black!0,backgroundcolor=black!1]
}{
    \end{mdframed}
  \end{originalfigure}
}

\theoremstyle{plain}

\theoremstyle{definition}

\def\d{{\rm d}}
\def\i{{\mathsf i}}
\DeclareMathOperator{\Tr}{Tr}
\DeclareMathOperator{\Det}{Det}

\DeclareMathOperator{\tr}{tr}
\def\Re{\mathop{\mathrm{Re}}}
\def\Im{\mathop{\mathrm{Im}}}
\def\diag{\mathop{\rm diag}\nolimits}
\def\sign{\mathop{\rm sign}\nolimits}

\usepackage{slashed}

\def\cD{{\cal D}}

\def\cK{{\cal K}}

\def\cO{{\cal O}}

\def\cR{{\cal R}}

\def\cT{{\cal T}}

\def\cV{{\cal V}}

\def\U{\mathrm{U}}

\def\beq#1\eeq{\begin{align}#1\end{align}}

\usepackage[all]{xy}

\begin{document}

\hypersetup{pageanchor=false}
\begin{titlepage}

\begin{flushright}
TU-1334
\end{flushright}

\vskip 3cm

\begin{center}

{\Large \bfseries Second-order formulation of chiral fermions\\ with a local boundary condition}

\vskip 1cm

Kazuya Yonekura~$^1$

\vskip 1cm

\begin{tabular}{ll}
$^1$Department of Physics, Tohoku University, Sendai 980-8578, Japan
\end{tabular}

\vskip 1cm

\end{center}

\noindent

Chiral fermions in $d$ dimensions arise naturally as boundary modes of a massive fermion in $d+1$ dimensions subject to a local boundary condition. When anomalies are absent, this construction can be used to define an intrinsically $d$-dimensional theory of chiral fermions. However, defining the path integral using the first-order Dirac operator with this boundary condition is technically inconvenient, even in the continuum. We present a formulation in terms of second-order differential operators that is completely equivalent to the original formulation in the continuum. Its natural lattice discretization generates Wilson's term and avoids fermion doubling. Our lattice formulation closely resembles the boundary-fermion construction of Shamir, and provides some insight into the case of a general Wilson parameter.

\end{titlepage}
\hypersetup{pageanchor=true}

\setcounter{tocdepth}{3}


\tableofcontents


\section{Introduction and summary}\label{sec:intro}

To define chiral fermions in $d$ dimensions, it is natural to use a theory in $d+1$ dimensions. Perturbative anomalies are described by anomaly inflow from Chern--Simons terms in $d+1$ dimensions~\cite{Callan:1984sa}. Global anomalies also admit a description in one higher dimension~\cite{Witten:1985xe,Witten:2015aba}. When perturbative anomalies cancel, the remaining global anomalies are characterized by bordism invariants of closed $(d+1)$-manifolds with the relevant gauge and spacetime structures~\cite{Kapustin:2014dxa,Freed:2016rqq,Yonekura:2018ufj}. A nonperturbative formulation of anomaly inflow for fermions, incorporating both perturbative and global anomalies, is given in~\cite{Witten:2019bou} using a massive fermion in $d+1$ dimensions with a local boundary condition. It applies to relativistic fermions in any dimension and with arbitrary symmetries. The anomaly-free condition is therefore also expressed in $d+1$ dimensions.

We begin by recalling the formulation of~\cite{Witten:2019bou} for a complex fermion, with no Majorana condition imposed. Let $Y$ be a Riemannian manifold of dimension $D=d+1$ with boundary $W=\partial Y$, and let $\Psi$ be a complex fermion on $Y$ of mass $m$, subject to the local boundary condition
\beq
\mathsf L:\qquad (1-\gamma^\tau)\Psi|_W=0, \label{eq:L-intro}
\eeq
where $\gamma^\tau$ is the gamma matrix along the outward normal. For $m<0$ this system has a massless chiral fermion localized along $W$.
We introduce a Pauli--Villars field of positive mass $M$, with the same boundary condition $\mathsf L$. The partition function is then given by
\beq
Z(Y,\mathsf L)=\frac{\Det(\cD_Y+\i m)}{\Det(\cD_Y+\i M)}, \label{eq:PV-ratio-intro}
\eeq
where $\cD_Y = \i \slashed{D}_Y$ is the Dirac operator on $Y$.
At distances large compared with $1/|m|$, this partition function becomes
\beq
Z(Y,\mathsf L)=|\Det \cD_W^+|\,\exp(-2\pi\i\eta_D). \label{eq:WY-intro}
\eeq
Here $\cD_W^+$ is the Dirac operator of $W$ acting on fermions of positive chirality, and $|\Det\cD_W^+|$ is the absolute value of the partition function of the chiral fermion on $W$. The factor $\exp(-2\pi\i\eta_D)$, where $\eta_D$ is the Atiyah--Patodi--Singer (APS) $\eta$-invariant~\cite{Atiyah:1975jf}, is the contribution of the bulk~\cite{Alvarez-Gaume:1984zst}.\footnote{We normalize $\eta_D$ as one half of the regulated spectral asymmetry of the bulk Dirac operator with APS boundary conditions, counting each zero mode with positive sign. This is half the normalization used in~\cite{Witten:2019bou}. It is the quantity denoted by $\xi$ in~\cite{Dai:1994kq}.}
Formula~\eqref{eq:WY-intro} was proposed in~\cite{Witten:2015aba} using the Dai--Freed theorem~\cite{Dai:1994kq} whose physical derivation is given in~\cite{Yonekura:2016wuc}. Then the derivation of the formula~\eqref{eq:WY-intro} itself was given in~\cite{Witten:2019bou}. For earlier work, see~\cite{Alvarez-Gaume:1985jck,DellaPietra:1986qd,Kaplan:1995pe}.

For an anomaly-free set of fermions, $\exp(-2\pi\i\eta_D)$ depends only on $W$ and the data on it, and \eqref{eq:WY-intro} can then be used as the definition of the chiral fermion partition function on $W$. Thus the $(d+1)$-dimensional path integral~\eqref{eq:WY-intro} determines the quantity corresponding to the path integral measure of the $d$-dimensional chiral fermions, including its phase.

To obtain a formulation with finitely many degrees of freedom, we seek a lattice regularization. Lattice gauge theory~\cite{Wilson:1974sk} gives a nonperturbative definition of gauge theories, but chiral gauge theories, such as the Standard Model, are notoriously hard to define. The basic obstruction is fermion doubling: a local lattice discretization of a first-order Dirac operator produces additional massless fermions, and they cannot be removed while keeping the chiral symmetry~\cite{Nielsen:1981hk}. For earlier work on this problem, see~\cite{Ginsparg:1981bj,Narayanan:1993zzh,Narayanan:1993ss,Narayanan:1993sk,Narayanan:1994gw,Neuberger:1997fp,Neuberger:1998wv,Hasenfratz:1998ri,Hasenfratz:1998jp,Luscher:1998pqa,Luscher:1998du,Luscher:1999un,Luscher:2000zd,Suzuki:2000ku,Kadoh:2007xb,Kikukawa:2017ngf,Luscher:2000hn,Kaplan:2009yg}. There has recently been renewed progress on this old problem~\cite{Craig:2026soq,Hsu:2026gmi,Witten:2026ezr} (see also~\cite{Liu:2026fla,Misumi:2026ham,Lu:2026vqh}). In particular, \cite{Witten:2026ezr} describes a lattice regularization, proposed also in~\cite{Hsu:2026gmi}, in which admissible lattice gauge fields are interpolated over the continuum and coupled to continuum chiral fermions (for earlier work, see~\cite{Gockeler:1992bi,tHooft:1994jaf,Hernandez:1995jc,Hernandez:1996rd}). Then the phase of the partition function is defined by $\exp(-2\pi\i\eta_D)$ using the interpolation.

To compute the phase $\exp(-2\pi\i\eta_D)$ without introducing fermion doublers, \cite{Witten:2026ezr} proposes using the eigenmodes of a second-order differential operator to evaluate $\eta_D$. This prescription uses the APS boundary condition, however, while the underlying theory in~\eqref{eq:WY-intro} is defined with the local condition $\mathsf L$ given by \eqref{eq:L-intro}. For a direct lattice formulation, it may be useful to combine the advantage of a second-order operator with a direct implementation of $\mathsf L$.

Even in the continuum, defining the path integral using the first-order Dirac operator with $\mathsf L$ is technically inconvenient. The operator $\cD_Y$ is not self-adjoint, because its adjoint obeys the opposite boundary condition, and its eigenvalues are not real. That alone would not be a problem. The problem is that its eigenvectors need not form a complete set. The usual definition of a fermion path integral, which expands the fields in eigenvectors, does not work in a straightforward way.

In this paper we define the fermion path integral with the boundary condition $\mathsf L$ directly, without passing through the APS condition. We look for a definition with the following properties: 
\begin{itemize}
  \item It should be completely equivalent to the formulation leading to \eqref{eq:WY-intro} in the continuum.
  \item It should allow an expansion in terms of eigenspaces of a self-adjoint operator.
  \item It should lead naturally to a lattice formulation without doublers. 
\end{itemize}
Our starting point is to rewrite the operator quotient appearing in the Pauli--Villars ratio~\eqref{eq:PV-ratio-intro} formally as a ratio of second-order operators,
\beq
\frac{\cD_Y+\i m}{\cD_Y+\i M}=\frac{(\cD_Y+\i m)(\cD_Y+\i M)^\dagger}{(\cD_Y+\i M)(\cD_Y+\i M)^\dagger}. \label{eq:rewrite-intro}
\eeq
The adjoint factor $(\cD_Y+\i M)^\dagger$ acts on spinors with the chirality condition opposite to $\mathsf L$. We introduce a field $\Phi$ with this condition and write the physical field as $\Psi=(\cD_Y+\i M)^\dagger\Phi$. This image must obey $\mathsf L$, which gives a second boundary condition on $\Phi$. 

With the above boundary conditions, the denominator is a positive self-adjoint operator of Laplace type, so its eigenvectors form a complete orthonormal set. Expanding $\Phi$ and the conjugate field in this set gives the physical fermion and its regulators a common space of fields. 

The second-order form is also what makes a lattice formulation natural. The numerator of \eqref{eq:rewrite-intro} has the differential expression
\beq
(M-m)\left(\slashed D_Y+\frac{\cD_Y^2}{M-m}+\frac{mM}{M-m}\right). \label{eq:mp-intro}
\eeq
The operator $\cD_Y^2=-D^\mu D_\mu + \cdots$ is of Laplace type. Discretizing its second-order part by a nearest-neighbor lattice Laplacian produces Wilson's term~\cite{Wilson:1975id}, which lifts the doublers. The mass $mM/(M-m)$ is negative and produces the chiral boundary fermion.

We discretize the second-order operators together with their boundary conditions to define the lattice path integral. For free fields, we identify a mass range in which an isolated boundary supports exactly one massless chiral fermion and no other massless mode. With a suitable choice of the mass parameters, our lattice formulation closely resembles the boundary-fermion construction of Shamir~\cite{Shamir:1993zy}.

In both the continuum formulation~\eqref{eq:rewrite-intro} and its lattice version, the denominator is positive and has a mass gap set by $M$. Its determinant has no phase, and its contribution to the effective action is expected to consist only of local terms in the large-$M$ limit. In the continuum, the equivalence with \eqref{eq:PV-ratio-intro} then implies that the phase of the numerator determinant reproduces $\exp(-2\pi\i\eta_D)$ in~\eqref{eq:WY-intro} in the large-mass limit. For the lattice formulation we expect the same after the continuum limit is also taken.

The paper is organized as follows. Section~\ref{sec:setup} reviews the massive fermion with the boundary condition $\mathsf L$ and fixes the notation. Section~\ref{sec:first-order} explains the problem of the first-order operator in the simplest example. Section~\ref{sec:continuum} gives the continuum formulation with second-order operators. Section~\ref{sec:lattice} gives the lattice formulation and discusses how the chiral fermion appears. Section~\ref{sec:discussion} summarizes the results and discusses the relation to other work.

\section{The massive fermion with the boundary condition \texorpdfstring{$\mathsf L$}{L}}\label{sec:setup}

We follow the setting of~\cite{Witten:2019bou,Witten:2026ezr}. Throughout this paper, we consider only complex fermions, with no Majorana condition imposed. In what follows, we refer to them simply as fermions. 

$Y$ denotes a compact Riemannian manifold of dimension $D=d+1$ with boundary $\partial Y = W$, together with any gauge field. We assume whatever structure (spin, pin, and so on) the fermion needs. Near $W$ the metric is a product, $\d s^2=\d\tau^2+\d s_W^2$, with $W$ at $\tau=0$ and $Y$ at $\tau<0$.\footnote{When $Y$ is oriented, we use the ``outward normal first'' convention for the boundary orientation. Thus $(\tau,x^1,\ldots,x^d)$ is positively oriented near $W$ when $(x^1,\ldots,x^d)$ is positively oriented on $W$. Our choice $\tau<0$ for the interior explains the use of intervals such as $[-\ell,0]$, with $\ell>0$, later in this paper. This choice is also convenient for the Hamiltonian interpretation in~\cite{Witten:2019bou}, where $\tau$ is Euclidean time and evolution proceeds toward the boundary at $\tau=0$. Unfortunately, this leads to somewhat unusual conventions.} We also take the gauge field to be independent of $\tau$ near $W$ and work in a gauge with no component along $\tau$ there.

The action is
\beq
I=-\int_Y \d^Dx\sqrt g\,\overline\Psi(\slashed D_Y+m)\Psi. \label{eq:action}
\eeq
We define
\beq
\cD_Y=\i\slashed D_Y, \qquad \langle\Phi',\Psi\rangle=\int_Y\d^Dx\sqrt g\,(\Phi')^\dagger\Psi.
\eeq
Here $\Phi'$ denotes an independent test spinor, and the prime does not denote differentiation.
With this convention, $\slashed D_Y+\mu=-\i(\cD_Y+\i\mu)$ for any mass $\mu$. 

Integration by parts gives
\beq
\langle\Phi',\cD_Y\Psi\rangle-\langle\cD_Y\Phi',\Psi\rangle=\i\int_W(\Phi')^\dagger\gamma^\tau\Psi, \label{eq:green}
\eeq
because $(\Phi')^\dagger(\i\gamma^\mu D_\mu\Psi)-(\i\gamma^\mu D_\mu\Phi')^\dagger\Psi=\i\nabla_\mu((\Phi')^\dagger\gamma^\mu\Psi)$. Here and below $\gamma^\tau$ is the gamma matrix along the outward normal, and the integral over $W$ uses the induced metric. On a manifold without boundary, \eqref{eq:green} says that $\cD_Y$ is self-adjoint.

Near the boundary,
\beq
\cD_Y=\i\gamma^\tau(\partial_\tau+\cD_W),\qquad \cD_W=\sum_{\mu\neq\tau}\gamma^\tau\gamma^\mu D_\mu. \label{eq:DW}
\eeq
The operator $\cD_W$ is self-adjoint on $W$ and anticommutes with $\gamma^\tau$. It is equivalent to the usual Dirac operator of $W$. We call $\gamma^\tau$ a chirality operator in any dimension $d$. We write 
\beq
P_\pm=\tfrac12(1\pm\gamma^\tau) 
\eeq
and denote the two local boundary conditions by
\beq
\begin{alignedat}{2}
&\mathsf L:\quad &&P_-\Psi|_W=0,\qquad \overline\Psi P_+|_W=0,\\
&\mathsf L':\quad &&P_+\Psi|_W=0,\qquad \overline\Psi P_-|_W=0.
\end{alignedat}\label{eq:LLprime}
\eeq
These conditions make the boundary term $\int_W\overline\Psi\gamma^\tau\delta\Psi$ in the variation of the action vanish. For $\mathsf L$, the column spinor $\overline\Psi{}^\dagger$ obeys $\mathsf L'$.

\paragraph{The adjoint obeys $\mathsf L'$.} If $\Psi$ obeys $\mathsf L$, then on $W$ we have $(\Phi')^\dagger\gamma^\tau\Psi=(\Phi')^\dagger\gamma^\tau P_+\Psi=(P_+\Phi')^\dagger\Psi$. So the boundary term in \eqref{eq:green} vanishes for every such $\Psi$ exactly when $\Phi'$ obeys $\mathsf L'$. Hence
\beq
(\cD_Y\ \text{with}\ \mathsf L)^\dagger=\cD_Y\ \text{with}\ \mathsf L'. \label{eq:adjoint}
\eeq
This makes precise the statement that $\mathsf L$ is not self-adjoint: the adjoint problem has the opposite chirality condition.

\paragraph{The sign of the mass.} For $\Psi$ obeying $\mathsf L$ we have $\Psi^\dagger\gamma^\tau\Psi=|\Psi|^2$ on $W$, so \eqref{eq:green} with $\Phi'=\Psi$ gives $\Im\langle\Psi,\cD_Y\Psi\rangle=\tfrac12\int_W|\Psi|^2$. We denote the $L^2$ norm on $Y$ by $\|\Psi\|^2=\langle\Psi,\Psi\rangle=\int_Y\d^Dx\sqrt g\,\Psi^\dagger\Psi$. Expanding the square, we find for real $\mu$
\beq
\|(\cD_Y+\i\mu)\Psi\|^2=\|\cD_Y\Psi\|^2+\mu^2\|\Psi\|^2+\mu\int_W|\Psi|^2\qquad(\Psi\ \text{obeys}\ \mathsf L). \label{eq:norm-identity}
\eeq
In the same way, $\|(\cD_Y-\i\mu)\Phi'\|^2=\|\cD_Y\Phi'\|^2+\mu^2\|\Phi'\|^2+\mu\int_W|\Phi'|^2$ if $\Phi'$ obeys $\mathsf L'$. For $\mu=M>0$ the right-hand sides are at least $M^2$ times the squared norm. Thus neither $\cD_Y+\i M$ with $\mathsf L$ nor its adjoint has a zero mode, and $\cD_Y+\i M$ with $\mathsf L$ is invertible.

For $\mu=m<0$ the boundary term in \eqref{eq:norm-identity} is negative and can cancel the other two. This is what allows the boundary mode
\beq
\Psi= \sqrt{2|m|}\, \chi\, e^{|m|\tau},\qquad (1-\gamma^\tau)\chi=0. \label{eq:edge-mode}
\eeq
On a half-cylinder $(-\infty,0]\times W$, the normalization gives $\|\Psi\|^2=\int_W|\chi|^2$. This mode has $\|\cD_Y\Psi\|^2=m^2\|\Psi\|^2+\int_W|\cD_W\chi|^2$ and $\int_W|\Psi|^2=2|m|\,\|\Psi\|^2$, so at $\mu=m$ the right-hand side of \eqref{eq:norm-identity} reduces to $\int_W|\cD_W\chi|^2$ and vanishes when $\cD_W\chi=0$. It is a massless fermion of positive chirality propagating along $W$. A Pauli--Villars field is given a positive mass so that it has no boundary mode.

\paragraph{The partition function.} For simplicity of presentation, we discuss the partition function, but correlation functions can be treated in the same way.

The integral over $\Psi$ and $\overline\Psi$ gives $\Det(\slashed D_Y+m)=\Det(-\i(\cD_Y+\i m))$. A Pauli--Villars field of mass $M>0$, with opposite statistics and the same boundary condition, divides it by the same expression at $m=M$. Formally, therefore,
\beq
Z(Y,\mathsf L)=\frac{\Det(\cD_Y+\i m)}{\Det(\cD_Y+\i M)}, \label{eq:Z-formal}
\eeq
the constant factors canceling in the ratio. A single Pauli--Villars field does not remove all ultraviolet divergences. Complete regularization requires several Pauli--Villars fields with different masses and statistics. We return to this in Section~\ref{sec:PV}.

\paragraph{The chiral fermion partition function.} If all perturbative and global anomalies of the boundary fermion $\chi$ cancel and appropriate local counterterms are added on $Y$, the partition function $Z(Y,\mathsf L)$ defines the partition function $Z_{\mathrm{chiral}}(W)$ of the chiral fermions on $W$ in the large-mass limit,
\beq
\lim_{\substack{m\to-\infty\\M\to+\infty}} Z(Y,\mathsf L)=Z_{\mathrm{chiral}}(W). \label{eq:chiral-limit}
\eeq
The result depends only on the boundary data and not on the choice of $Y$ or the extension of these data into it.

\paragraph{A fixed reference gauge field.} Let $W_0$ be a closed $d$-dimensional manifold and take $Y=[-\ell,0]\times W_0$, with $\ell>0$. Consider a fixed topological class of gauge bundles on $W_0$. On the boundary at $\tau=-\ell$, we choose a fixed representative gauge field $A_*$ of the given topology, and on the boundary at $\tau=0$ take a general gauge field $A$ in the same topological class. Extend them into $Y$ by any smooth interpolation that is independent of $\tau$ near $\tau=-\ell$ and $\tau=0$. Under the same assumption of anomaly cancellation, \eqref{eq:chiral-limit} gives
\beq
\lim_{\substack{m\to-\infty\\M\to+\infty}} Z(Y,\mathsf L)
=Z_{\mathrm{chiral}}(W_0,A)\,\overline{Z_{\mathrm{chiral}}(W_0,A_*)}. \label{eq:cylinder-factorization}
\eeq
The complex conjugation accounts for the chirality at $\tau=-\ell$ being opposite to that at $\tau=0$. For a choice of $A_*$ with a nonzero partition function, the contribution from the boundary at $\tau=-\ell$ is a nonzero constant independent of $A$. Thus the cylinder defines the chiral fermion path integral at $A$ up to this constant, independently of the bulk interpolation.\footnote{If zero modes force the partition function to vanish throughout a topological sector, one instead uses correlation functions with fermion insertions that absorb the zero modes, choosing a nonzero reference correlation function.}

The absolute value of the reference partition function $Z_{\mathrm{chiral}}(W_0,A_*)$ can be defined without difficulty. Its phase can be chosen consistently using the bordism construction described in~\cite{Witten:2019bou,Witten:2026ezr}, to which we refer for details.

\section{The first-order operator in the simplest example}\label{sec:first-order}

The usual way to give \eqref{eq:Z-formal} a meaning is to use the eigenvectors of $\cD_Y$. When $\cD_Y$ is self-adjoint, as on a closed manifold or with APS boundary conditions, they form a complete orthonormal set. Expanding $\Psi$ and $\overline\Psi$ in them makes the action diagonal, and the fermionic Gaussian integral becomes a product of eigenvalues. Then the ratio \eqref{eq:Z-formal} becomes $\prod_k(\lambda_k+\i m)/(\lambda_k+\i M)$. 

With $\mathsf L$, the operator $\cD_Y$ is not self-adjoint by \eqref{eq:adjoint}, and its eigenvalues are not real. This alone would not prevent the construction. If the eigenvectors of $\cD_Y$ formed a complete set, one could expand $\Psi$ in them and $\overline\Psi$ in the dual set. Since $\cD_Y+\i m$ and $\cD_Y+\i M$ have the same eigenvectors, the Jacobians of these non-orthogonal bases would cancel in the ratio. What can fail is completeness. In the simplest example it fails completely.

We now turn to a concrete example. Take $D=1$ and $Y=[-\ell,0]$. Then $W$ consists of the two points $\tau=0$ and $\tau=-\ell$, a spinor has one component, and we may take $\gamma^\tau=1$ along the coordinate $\tau$. At $\tau=0$ the outward normal is $+\partial_\tau$, and $\mathsf L$ leaves $\Psi$ unconstrained and requires $\overline\Psi(0)=0$. At $\tau=-\ell$ the outward normal is $-\partial_\tau$, the gamma matrix along it is $-1$, and $\mathsf L$ requires $\Psi(-\ell)=0$ and leaves $\overline\Psi$ unconstrained. So the operator acting on $\Psi$ is
\beq
\cD_Y=\i\frac{\d}{\d\tau}\qquad\text{on functions on }[-\ell,0]\text{ with }\Psi(-\ell)=0. \label{eq:1d-operator}
\eeq
The same problem arises in any dimension. On a slab $[-\ell,0]\times W_0$ with a background that does not depend on $\tau$, put $\Psi=u(\tau)\chi$, where $\chi$ is a zero mode of the Dirac operator $\cD_{W_0}$ with $\gamma^\tau\chi=\chi$. Then $\cD_Y\Psi=(\i u')\chi$, and $\mathsf L$ requires $u(-\ell)=0$ and nothing at $\tau=0$. Such zero modes, or their mirror images with $\gamma^\tau\chi=-\chi$, are present whenever $\cD_{W_0}$ has a nonzero index, for instance when $W_0$ is a two-torus with magnetic flux, and the index guarantees that they persist when the background is deformed.

The eigenvalue equation $\i\Psi'=\lambda\Psi$ has the solutions $\Psi=c\,e^{-\i\lambda\tau}$, and $\Psi(-\ell)=0$ forces $c=0$.
Therefore, the operator \eqref{eq:1d-operator} has no eigenvalue at all.\footnote{A similar phenomenon occurs in quantum mechanics. For a particle in an infinite square well $0<x<\ell$, the momentum operator $p=-\i\hbar\,\d/\d x$ has no eigenfunctions satisfying $\psi(0)=\psi(\ell)=0$. Its eigenfunctions would be exponentials, and either endpoint condition forces them to vanish identically. With this Dirichlet domain, $p$ is symmetric but not self-adjoint, since its adjoint imposes no endpoint conditions. Thus $p$ with this domain is not an observable in the usual self-adjoint sense.}

The consequences are immediate. The eigenvectors of $\cD_Y$ span nothing, so they cannot be used to expand the fields or to define a cutoff, and there is no analog of the derivation of $\prod_k(\lambda_k+\i m)/(\lambda_k+\i M)$. One could still write down that product, but here it is the empty product, equal to $1$ for every $m$ and $M$.

The limitation of a first-order eigenvalue product for $\mathsf L$ was already pointed out in~\cite{Ivanov:2022jor}. That work shows that $\mathsf L$ is elliptic but fails a stronger condition, strong ellipticity, on which its heat-kernel methods rely. The interval above exhibits the obstruction directly, without requiring the general theory of elliptic boundary problems.

Nevertheless $\cD_Y+\i\mu$ is invertible. Its inverse is
\beq
\big((\cD_Y+\i\mu)^{-1}f\big)(\tau)=-\i\int_{-\ell}^{\tau}e^{-\mu(\tau-\tau')}f(\tau')\,\d\tau'. \label{eq:volterra}
\eeq
The propagator carries the field forward in $\tau$ only. At $\mu=m$, it depends nontrivially on the physical mass $m$. For $m<0$ it grows toward $\tau=0$, reflecting the boundary-localized mode of the half-line. (On the finite interval the condition at $\tau=-\ell$ prevents an exact zero mode.)

The propagator is a nonzero compact operator, but it has no eigenvalue either: an eigenvector of \eqref{eq:volterra} with a nonzero eigenvalue would be an eigenvector of $\cD_Y$, and the propagator is injective, so zero is not an eigenvalue.\footnote{The spectrum of the propagator $(\cD_Y+\i\mu)^{-1}$ is $\{0\}$. Zero belongs to the spectrum because the inverse of the propagator is the unbounded operator $\cD_Y+\i\mu$, but it is not an eigenvalue. The spectrum of $\cD_Y$ itself is empty.}

For a general $Y$ in the setting of Section~\ref{sec:setup}, it is shown in~\cite{Witten:2026ezr} that $\cD_Y$ with $\mathsf L$ has a compact inverse, so that its spectrum is discrete, consisting of isolated eigenvalues of finite multiplicity. The interval example is consistent with this result, since a discrete spectrum may be empty. Without a complete eigenbasis, the first-order formulation is technically inconvenient for defining the path integral. The second-order reformulation below supplies a convenient basis for the same theory.

For the Pauli--Villars field, the propagator and its kernel are
\beq
\begin{aligned}
R=(\cD_Y+\i M)^{-1},\qquad 
R(\tau,\tau')=-\i e^{-M(\tau-\tau')}\theta(\tau-\tau'),
\end{aligned} \label{eq:pv-propagator-interval}
\eeq
where $\theta$ is the step function.
Formally $\cD_Y+\i m=(\cD_Y+\i M)[1-\i(M-m)R]$. So the logarithm of \eqref{eq:Z-formal} is a sum of one-loop diagrams, $-\sum_{q\geq1}\frac{1}{q}[\i(M-m)]^q\,\Tr R^q$. Since $R$ carries the field forward only ($R(\tau,\tau')=0$ for $\tau < \tau'$), the kernel of $R^q$ vanishes at coincident points $\tau=\tau'$ for $q\geq2$, and so do these loops, as an empty spectrum requires. The loop with one propagator is the integral of $R(\tau,\tau)$, where the kernel jumps from $-\i$ to $0$ as $\tau'$ passes $\tau$. Its value depends on how this coincident point is treated, and the spectrum of $\cD_Y$ does not fix that choice.

For example, approximate $\theta(x)$ by truncating its Fourier expansion to finitely many modes. Since $\theta(x)-1/2$ is odd, this approximation consists of the constant $1/2$ and a finite sum of sine modes. Each sine mode vanishes at $x=0$, so the approximation takes the value $1/2$ there at every cutoff. This defines a prescription with $\theta(0)=1/2$ and hence $R(\tau,\tau)=-\i/2$.
With this prescription, the single physical fermion and its Pauli--Villars field give
\beq
\begin{aligned}
\log Z(Y,\mathsf L)
=-\i(M-m)\int_{-\ell}^{0}R(\tau,\tau)\,\d\tau
=-\frac{(M-m)\ell}{2},\qquad
Z(Y,\mathsf L)=e^{-(M-m)\ell/2}.
\end{aligned}\label{eq:interval-midpoint-partition}
\eeq

The additional Pauli--Villars fields (Section~\ref{sec:PV}) cancel this loop. In this example the fully regularized partition function is then $1$, as is the empty product.\footnote{After complete Pauli--Villars regularization, $f(R)-1$ in Section~\ref{sec:PV} has summable singular values. Thus $\Det f(R)$ equals the product of the eigenvalues of $f(R)$ with algebraic multiplicity, even without a complete eigenbasis (Lidskii's theorem).} There is thus no fundamental problem. The only technical inconvenience is that the first-order operator does not provide a basis for the path integral. The information that a construction needs sits instead in the propagator, which exists although the eigenvectors do not.

\section{Continuum formulation with second-order operators}\label{sec:continuum}

This section defines the regulated path integral with the boundary condition $\mathsf L$ using the rewriting \eqref{eq:rewrite-intro}. 

\subsection{The second-order operators and their boundary conditions}\label{sec:second-order}

For a real mass $\mu$, let
\beq
\cT_\mu=\cD_Y+\i\mu\qquad\text{acting on spinors that obey }\mathsf L. \label{eq:T}
\eeq
By \eqref{eq:adjoint}, its adjoint is $\cT_\mu^\dagger=\cD_Y-\i\mu$ acting on spinors that obey $\mathsf L'$. In this notation \eqref{eq:Z-formal} is formally $\Det \cT_m/\Det \cT_M$. Following \eqref{eq:rewrite-intro}, we define
\beq
\cK_\mu=\cT_\mu \cT_M^\dagger. \label{eq:K-def}
\eeq
Since the masses are constants, its differential expression is
\beq
\cK_\mu=\cD_Y^2-\i(M-\mu)\cD_Y+\mu M. \label{eq:K}
\eeq
The numerator of \eqref{eq:rewrite-intro} is $\cK_m$ and the denominator is $\cK_M=\cD_Y^2+M^2$.

An operator includes its boundary conditions as well as its differential expression. The product \eqref{eq:K-def} acts on a spinor $\Phi$ that obeys $\mathsf L'$ and whose image $\cT_M^\dagger\Phi$ obeys $\mathsf L$. At the boundary, the term involving derivatives along $W$ satisfies $P_-\gamma^\tau\cD_W\Phi=\gamma^\tau\cD_WP_+\Phi=0$, since $\cD_W$ anticommutes with $\gamma^\tau$. Using \eqref{eq:DW} and $P_-\gamma^\tau=-P_-$, we find $P_-\cT_M^\dagger\Phi|_W=-\i(\partial_\tau+M)P_-\Phi|_W$. The domain of $\cK_\mu$ therefore has the boundary conditions
\beq
P_+\Phi|_W=0,\qquad (\partial_\tau+M)P_-\Phi|_W=0. \label{eq:K-bc}
\eeq
The first is $\mathsf L'$. The second sets a linear combination of $P_-\Phi$ and its outward normal derivative $\partial_\tau P_-\Phi$ to zero at $W$. Such a condition is called a Robin boundary condition. The coefficient multiplying $P_-\Phi$ is $M$ for every $\mu$, because the factor that acts first is always $\cT_M^\dagger$.

The numerator $\cK_m$ and the denominator $\cK_M$ are obtained from \eqref{eq:K} by setting $\mu=m$ and $\mu=M$, respectively. Both act on spinors obeying the boundary conditions \eqref{eq:K-bc}, which involve the regulator mass $M$ but not $m$.
The physical mass $m$ enters only through the coefficients of the differential operator. Section~\ref{sec:measure} defines the determinant ratio on this common domain.

The denominator is positive. For $\Phi$ obeying \eqref{eq:K-bc}, the version of \eqref{eq:norm-identity} for $\mathsf L'$ gives
\beq
\langle\Phi,\cK_M\Phi\rangle=\|\cT_M^\dagger\Phi\|^2=\|\cD_Y\Phi\|^2+M^2\|\Phi\|^2+M\int_W|\Phi|^2\ \geq\ M^2\|\Phi\|^2. \label{eq:KM-positive}
\eeq
Thus $\cK_M$ is a self-adjoint operator of Laplace type with strictly positive spectrum. Its eigenvectors form a complete orthonormal set,
\beq
\cK_Mu_j=s_j^2u_j,\qquad \langle u_j,u_k\rangle=\delta_{jk},\qquad s_j\geq M. \label{eq:uj}
\eeq
The numerator $\cK_m$ is not self-adjoint, and we will not need its eigenvectors.

\subsection{The space of fields and the measure}\label{sec:measure}

The action of a field of mass $\mu$ is, by \eqref{eq:action},
\beq
I_\mu=-\int_Y\d^Dx\sqrt g\,\overline\Psi(\slashed D_Y+\mu)\Psi=\i\,\langle\overline\Psi{}^\dagger,\cT_\mu\Psi\rangle.
\eeq
The fields $\Psi$ and $\overline\Psi{}^\dagger$ obey $\mathsf L$ and $\mathsf L'$, respectively, so the bilinear pairs two different linear spaces. To use fields obeying $\mathsf L'$ on both sides, we make the change of variables
\beq
\Psi=\cT_M^\dagger\Phi,\qquad \Phi\ \text{obeys \eqref{eq:K-bc}}. \label{eq:change}
\eeq
Since $\cT_M^\dagger$ is invertible for $M>0$, this change of variables is one-to-one. It does not depend on $\mu$, so its Jacobian is the same for the physical fermion and for every regulator field, and it cancels in the ratio. Using $\cT_\mu \cT_M^\dagger=\cK_\mu$, the action becomes
\beq
I_\mu=\i\,\langle\overline\Psi{}^\dagger,\cK_\mu\Phi\rangle. \label{eq:action-K}
\eeq
Both $\Phi$ and $\overline\Psi{}^\dagger$ obey $\mathsf L'$, and $\Phi$ obeys in addition the Robin condition of \eqref{eq:K-bc}.

We expand both fields in the eigenvectors \eqref{eq:uj} of the denominator,
\beq
\Phi=\sum_ja_ju_j,\qquad \overline\Psi{}^\dagger=\sum_jb_ju_j, \label{eq:expansion}
\eeq
and define the measure as the product of the integrals over the independent Grassmann coefficients $a_j$ and $\overline b_j$. For $\Phi$ this is the natural expansion, since every $u_j$ obeys the boundary conditions \eqref{eq:K-bc} of $\Phi$, and these are what make $\Psi=\cT_M^\dagger\Phi$ obey $\mathsf L$. Each $u_j$ also obeys $\mathsf L'$, as $\overline\Psi{}^\dagger$ must.

Using one set for both fields makes the matrices below square, with a cutoff common to both fields. Defining a fermion measure through the eigenmodes of a second-order operator is familiar (see, for example,~\cite{Fujikawa:1983bg}). What matters here is that the same eigenmodes are used for every field.

As a provisional step, we keep the modes with $s_j$ below a common bound and call their number $N$. We denote their span by $U_N$, meaning the $N$-dimensional space of all linear combinations of the retained eigenvectors $u_1,\ldots,u_N$. This sharp cutoff is only an intermediate device for making the Gaussian integral finite-dimensional. Section~\ref{sec:PV} removes it after introducing additional Pauli--Villars fields, while Section~\ref{sec:lattice} presents an alternative lattice regularization. At finite $N$, the Pauli--Villars ratio becomes
\beq
Z_N(m;M)=\frac{\det \cK_{m,N}}{\det \cK_{M,N}},\qquad (\cK_{\mu,N})_{jk}=\langle u_j,\cK_\mu u_k\rangle\quad(j,k\leq N). \label{eq:ZN}
\eeq
The denominator matrix is $\cK_{M,N}=\diag(s_1^2,\ldots,s_N^2)$. Each $u_k$ obeys \eqref{eq:K-bc}, which does not involve $\mu$. This is the precise meaning of \eqref{eq:rewrite-intro}: a ratio of determinants of the second-order numerator and denominator on one common space.

Since the denominator $\det\cK_{M,N}$ is positive, the phase of $Z_N(m;M)$ comes entirely from the numerator $\det\cK_{m,N}$.

We record a few formulas for later use.
We use the Lichnerowicz formula $\cD_Y^2=-D^\nu D_\nu+\cR$ in \eqref{eq:K}, where $\cR=-\tfrac12\gamma^{\nu\lambda}[D_\nu,D_\lambda]$ and $\gamma^{\nu\lambda}=\tfrac12[\gamma^\nu,\gamma^\lambda]$. The term $\cR$ contains one quarter of the scalar curvature and the coupling of the spin to the gauge field strength.

For $\Phi'$ obeying $\mathsf L'$ and $\Phi$ obeying \eqref{eq:K-bc}, we integrate the covariant Laplacian in $\langle\Phi',\cK_\mu\Phi\rangle$ by parts. Its boundary term is $-\int_W(\Phi')^\dagger\partial_\tau\Phi$, which becomes $M\int_W(\Phi')^\dagger\Phi$ by $P_+\Phi'=0$ and the Robin condition. We obtain
\beq
\begin{aligned}
\langle\Phi',\cK_\mu\Phi\rangle
&=\langle \cT_\mu^\dagger\Phi',\cT_M^\dagger\Phi\rangle\\
&=\int_Y\d^Dx\sqrt g\Big[(D^\nu\Phi')^\dagger D_\nu\Phi+(\Phi')^\dagger\cR\,\Phi+\mu M\,(\Phi')^\dagger\Phi\Big]\\
&\quad -\i(M-\mu)\langle\Phi',\cD_Y\Phi\rangle+M\int_W(\Phi')^\dagger\Phi.
\end{aligned}\label{eq:local-form}
\eeq
The equality between $\langle \cT_\mu^\dagger\Phi',\cT_M^\dagger\Phi\rangle$ and the integral expression in \eqref{eq:local-form} requires only $\mathsf L'$ for both $\Phi'$ and $\Phi$.

\subsection{Removing the cutoff}\label{sec:PV}

To see whether \eqref{eq:ZN} has a limit as $N\to\infty$, write it in terms of the regulator propagator $R=\cT_M^{-1}$. Since $\cT_\mu=\cT_M-\i(M-\mu)$, the quotient with the denominator on the right is
\beq
\cK_\mu\cK_M^{-1}=\cT_\mu \cT_M^{-1}=1-\i(M-\mu)R. \label{eq:K-quotient}
\eeq
Recall that $U_N$ is spanned by the retained eigenvectors $u_1,\ldots,u_N$ of $\cK_M$, with $\cK_Mu_j=s_j^2u_j$. Thus $\cK_M$ maps $U_N$ into itself, and projecting \eqref{eq:K-quotient} onto $U_N$ and taking determinants gives
\beq
Z_N(\mu;M)=\det\big[1-\i(M-\mu)R_N\big],\qquad (R_N)_{jk}=\langle u_j,Ru_k\rangle\quad (j,k\leq N). \label{eq:ZN-R}
\eeq
This is the cutoff version of \eqref{eq:K-quotient}. Its logarithm is a sum of one-loop diagrams,
\beq
\log Z_N(\mu;M)=-\sum_{q\geq1}\frac{1}{q}\big[\i(M-\mu)\big]^q\,\tr R_N^q, \label{eq:loops}
\eeq
in which the term of order $q$ is the loop with $q$ regulator propagators and $q$ insertions of the mass difference. The size of these loops is controlled by the singular values of $R$, defined as the nonnegative square roots of the eigenvalues of $R^\dagger R$. In this case,
\beq
R^\dagger R=(\cT_M\cT_M^\dagger)^{-1}=\cK_M^{-1}, \label{eq:R-singular-values}
\eeq
so they are the numbers $1/s_j$. At large local momentum $p$, the leading term of $\cK_M$ is $p^2$, so the singular values of $R$ scale as $1/|p|$. The matrix $R_N$ represents $R$ followed by the orthogonal projection onto $U_N$, so its singular values are no larger than those of $R$.

High-energy states are counted by the phase-space measure $\d^Dx\,\d^Dp/(2\pi)^D$ per internal component, as in quantum mechanics. The ultraviolet convergence of a loop with $q$ propagators is therefore controlled by
\beq
\int_{|p|>p_0}\frac{\d^Dp}{(2\pi)^D}\frac{1}{|p|^q}
\ \propto\ \int_{p_0}^{\infty}\d p\,p^{D-1-q}, \label{eq:loop-power-counting}
\eeq
where $p_0$ is a fixed large momentum scale. This integral converges for $q>D$, so such loops converge absolutely, while those with $q\leq D$ can diverge. This is the familiar statement that a single Pauli--Villars field does not regularize everything.

We now include several Pauli--Villars fields. Let the fields have masses $\mu_j$ and integer weights $c_j$, $j=0,\ldots,n$, where $\mu_0=m$ and $c_0=1$ describe the physical fermion, $c_j>0$ counts anticommuting and $c_j<0$ commuting regulator fields, all regulator masses are positive, and
\beq
\sum_{j=0}^nc_j\,\mu_j^{\,q}=0\qquad(q=0,1,\ldots,n-1),\qquad n>D. \label{eq:PV-moments}
\eeq
For $m<0<M$, an explicit choice satisfying \eqref{eq:PV-moments} is $\mu_j=m+j(M-m)$ and $c_j=(-1)^j\binom{n}{j}$, so that $\mu_1=M$.\footnote{Apply $(x\partial_x)^r$ to the binomial identity $(1-x)^n=\sum_{j=0}^n(-1)^j\binom{n}{j}x^j$ and set $x=1$. For $0\leq r<n$, the left-hand side vanishes because all derivatives of order less than $n$ vanish at $x=1$, giving $\sum_{j=0}^nc_jj^r=0$. Since $\mu_j^q=[m+j(M-m)]^q$ is a polynomial in $j$ of degree $q<n$, it follows that $\sum_{j=0}^nc_j\mu_j^q=0$.} The additional regulator masses vary with $m$ to maintain the cancellations, but they enter only the action coefficients. All fields use the same space $U_N$, with the same boundary conditions \eqref{eq:K-bc}, and the same right factor $\cT_M^\dagger$ in \eqref{eq:K-def}.

For a regulator field of mass $\mu>0$, \eqref{eq:green} gives
\beq
\Re\,\langle\Phi,\cK_\mu\Phi\rangle=\|\cD_Y\Phi\|^2+\mu M\|\Phi\|^2+\frac{\mu+M}{2}\int_W|\Phi|^2>0\qquad(\Phi\neq0). \label{eq:Re-K}
\eeq
Thus the Gaussian integral of $\exp(-a^\dagger\cK_{\mu,N}a)$ over the commuting coefficients $a\in\mathbb C^N$ converges and is proportional to $1/\det\cK_{\mu,N}$.

The regulated partition function at the cutoff is then
\beq
Z_{{\rm PV},N}=\prod_{j=0}^nZ_N(\mu_j;M)^{c_j}=\det f(R_N),\qquad f(z)=\prod_{j=0}^n\big[1-\i(M-\mu_j)z\big]^{c_j}. \label{eq:ZPV-N}
\eeq
All factors are functions of the one matrix $R_N$, because all fields share the same space and the same right factor $\cT_M^\dagger$. In the loop expansion of $\log Z_{{\rm PV},N}$ the coefficient of $\tr R_N^q$ is proportional to $\sum_jc_j(M-\mu_j)^q$, which vanishes for $q<n$ by \eqref{eq:PV-moments}. Equivalently, $f(z)=1+O(z^n)$. So the loops with fewer than $n$ propagators, which include all loops that can diverge, cancel between the physical fermion and the regulators at every value of the cutoff. The remaining loops converge absolutely, and $Z_{{\rm PV},N}$ has a limit:
\beq
Z_{\rm PV}(Y,\mathsf L)=\lim_{N\to\infty}Z_{{\rm PV},N}=\det f(R). \label{eq:ZPV}
\eeq
The loop expansion is used only to see which terms cancel.

Once the cutoff is removed, the result is the determinant of an operator built from the propagator $R$ with the boundary condition $\mathsf L$. The eigenvectors of $\cK_M$ serve to construct it, as a common basis and a common cutoff, but they do not appear in the answer. The cutoff matters again for a single physical/Pauli--Villars pair, whose one-propagator loop needs a prescription.

In the interval example of Section~\ref{sec:first-order}, only $\tr R$ depends on the coincident-point prescription, while $\tr R^q=0$ for $q\geq2$. Canceling the $\tr R$ term with additional Pauli--Villars fields therefore leaves a result determined entirely by the propagator $R$, which is a compact operator.

Local counterterms must be chosen so that the renormalized bulk vacuum energy and the coefficients of local bulk terms such as the gauge kinetic term vanish. Otherwise the partition function would retain a dependence on $Y$ even in the anomaly-free large-mass limit \eqref{eq:chiral-limit}.

\subsection{Remarks on the choice of functions}\label{sec:sine-cutoff}

We have described the preceding construction in some detail, although many of its steps may seem elementary. To explain why this care is needed, we first examine a choice of functions that gives an incorrect determinant phase. We then use this failure to identify what a suitable approximation must preserve and explain how the second-order formulation provides a criterion for choosing functions. This also clarifies why the Robin condition need not be imposed on the basis functions.

\paragraph{Failure of a naive approach.}
As an example, take $Y=[-\ell,0]$, $\gamma^\tau=1$ along the coordinate $\tau$, and a constant $\U(1)$ gauge field,
\beq
A_\tau=\frac{\theta}{\ell},\qquad
D_\tau=\partial_\tau-\i\frac{\theta}{\ell},\qquad
\cT_\mu=\i D_\tau+\i\mu.
\label{eq:sine-background}
\eeq
As in Section~\ref{sec:first-order}, the boundary conditions require $\Psi(-\ell)=0$ and $\overline\Psi(0)=0$, while leaving $\Psi(0)$ and $\overline\Psi(-\ell)$ free. The gauge transformation with phase $e^{\i\theta(\tau+\ell)/\ell}$ removes $A_\tau$ and preserves these boundary conditions. Thus the continuum partition function has the same positive real value as \eqref{eq:interval-midpoint-partition} (up to the choice of a real value $\i R(\tau,\tau)$). In particular, its phase is $1$ for every $\theta$.

The first-order operator $\cT_\mu$ with $\mathsf L$ has no eigenvectors in this example as in Section~\ref{sec:first-order}, so it does not provide an eigenfunction expansion to guide our choice of functions for the path integral.

We therefore try a naive ansatz in which $\Psi$ and $\overline\Psi$ are expanded in the same set of functions. Since $\Psi(-\ell)=0$ and $\overline\Psi(0)=0$, each function in this common set must vanish at both endpoints. We choose the sine functions, which have this property and form a complete orthonormal basis of $L^2([-\ell,0])$,
\beq
\begin{aligned}
v_j(\tau)&=\sqrt{\frac2\ell}\sin\frac{j\pi(\tau+\ell)}\ell,
\qquad j=1,2,\ldots.
\end{aligned}
\label{eq:sine-basis}
\eeq
We introduce a cutoff by retaining the first $N$ functions and expanding both fields in their span. The gauge transformation above does not preserve this space, so it cannot be used to remove $A_\tau$ at fixed $N$. The question is whether we obtain a sensible answer in the limit $N \to \infty$.

Define the matrix $B_N$ by $(B_N)_{jk}=\ell\langle v_j,\partial_\tau v_k\rangle$. It is real because the basis functions are real, and antisymmetric by integration by parts since they vanish at both endpoints.
The projected first-order operator is $\cT_{\mu,N}=\i[B_N+(\mu\ell-\i\theta)\mathbf1_N]/\ell$, so its determinant ratio is
\beq
Z_N^{(1)}(\theta)
=\frac{\det[B_N+(m\ell-\i\theta)\mathbf1_N]}
{\det[B_N+(M\ell-\i\theta)\mathbf1_N]}.
\label{eq:sine-first-ratio}
\eeq
Since $B_N$ is real and antisymmetric, its nonzero eigenvalues occur in pairs $\pm\i\omega_{r,N}$, where $\omega_{r,N}>0$. For odd $N=2n+1$ there is also a zero eigenvalue. At $\theta=0$, each nonzero pair contributes a positive factor, while the zero eigenvalue contributes $m/M$. Thus
\beq
Z_{2n+1}^{(1)}(0)=\frac mM\prod_{r=1}^n
\frac{\omega_{r,2n+1}^2+(m\ell)^2}
{\omega_{r,2n+1}^2+(M\ell)^2}<0
\qquad(m<0<M).
\label{eq:sine-odd-failure}
\eeq
Odd cutoffs therefore give the wrong phase even at zero gauge field, for every $n$. Taking $n\to\infty$ cannot repair this failure. We henceforth restrict attention to even cutoffs $N=2n$.

For even $N$, pairing the eigenvalues in \eqref{eq:sine-first-ratio} gives
\beq
Z_{2n}^{(1)}(\theta)=\prod_{r=1}^n
\frac{\omega_{r,2n}^2+(m\ell-\i\theta)^2}
{\omega_{r,2n}^2+(M\ell-\i\theta)^2}.
\label{eq:sine-even-product}
\eeq
To test whether the phase is correct, it is enough to consider its derivative at $\theta=0$. Differentiating the logarithm of \eqref{eq:sine-even-product} gives
\beq
\begin{aligned}
\left.\partial_\theta\arg Z_{2n}^{(1)}(\theta)\right|_{\theta=0}
&=\left.\Im\,\partial_\theta\log Z_{2n}^{(1)}(\theta)\right|_{\theta=0}\\
&=2\ell\sum_{r=1}^n\left[
\frac{|m|}{\omega_{r,2n}^2+m^2\ell^2}
+\frac{M}{\omega_{r,2n}^2+M^2\ell^2}\right]>0.
\end{aligned}
\label{eq:sine-first-phase-response}
\eeq
Every term is positive for $m<0<M$, so there can be no cancellation.

When $|m|\ell,M\ell\gg1$, we can estimate \eqref{eq:sine-first-phase-response} by assuming the high-energy behavior $\omega_{r,2n}\sim r$. We take $n\gg|m|\ell,M\ell$ so that the relevant frequencies lie well below the cutoff. There are $O(|m|\ell)$ modes with $\omega_{r,2n}\lesssim|m|\ell$, each contributing $O(1/(|m|\ell))$ to the $|m|$ term in \eqref{eq:sine-first-phase-response}. Their sum is therefore of order one, and the convergent tail at $\omega_{r,2n}\gtrsim|m|\ell$ has the same scaling. The same estimate applies to the $M$ term.

We therefore expect the leading behavior of the phase derivative to be nonzero and independent of $|m|$ and $M$. The correct phase is independent of $\theta$, so its derivative is zero. Even $N$ therefore fail this test as well.

\paragraph{Choosing functions for the path integral.} In the example above, the sine functions $v_j$ form a complete orthonormal basis of $L^2([-\ell,0])$. This means that any allowed field $\Psi$ can be approximated by linear combinations $\Psi_N$ of $v_1,\ldots,v_N$ such that $\|\Psi-\Psi_N\|\to0$ as $N\to\infty$. The same holds for $\overline\Psi$. To approximate the first-order path integral, we must also approximate the quadratic form
\beq
\int_{-\ell}^{0}\d\tau\,\overline\Psi \cT_\mu\Psi
=\i\int_{-\ell}^{0}\d\tau\,\overline\Psi
\left(\partial_\tau+\mu-\i\frac{\theta}{\ell}\right)\Psi.
\label{eq:sine-first-quadratic-form}
\eeq
Approximating this expression requires control of the derivative as well. A small $L^2$ error in the fields alone does not provide that control.

The original boundary conditions allow $\Psi(0)$ and $\overline\Psi(-\ell)$ to be nonzero, but every sine expansion sets $\Psi_N(0)=0$ and $\overline\Psi_N(-\ell)=0$. For a smooth allowed profile $\Psi$ with $\Psi(0)\ne0$, its difference from any sine expansion obeys by the Schwarz inequality
\beq
\begin{aligned}
\big\|\partial_\tau(\Psi-\Psi_N)\big\|^2
\geq\frac1\ell\left|\int_{-\ell}^{0}\d\tau\,
\partial_\tau(\Psi-\Psi_N)\right|^2
=\frac{|\Psi(0)|^2}{\ell}.
\end{aligned}
\label{eq:sine-endpoint-error}
\eeq
Thus the field can be approximated arbitrarily well in $L^2$ while the error in its derivative cannot tend to zero. The same problem occurs for $\overline\Psi$ at $\tau=-\ell$. We therefore need a norm that controls derivatives as well as the fields themselves. The first-order form in \eqref{eq:sine-first-quadratic-form} is not positive definite and cannot itself define such a norm.

To obtain a positive form that includes derivatives, we can take the squared $L^2$ norm after applying a first-order operator. We use $q_M(\Phi',\Phi)=\langle \cT_M^\dagger\Phi',\cT_M^\dagger\Phi\rangle$ for spinors obeying $\mathsf L'$. The $\mathsf L'$ version of \eqref{eq:norm-identity} gives (without the Robin condition)
\beq
\begin{aligned}
q_M(\Phi,\Phi)&=\|\cT_M^\dagger\Phi\|^2\\
&=\|\cD_Y\Phi\|^2+M^2\|\Phi\|^2+M\int_W|\Phi|^2.
\end{aligned}
\label{eq:approximation-norm}
\eeq
Since $M>0$, $q_M(\Phi,\Phi)^{1/2}$ is a norm that controls the field, its derivatives, and its boundary values. The form $q_M$ is associated with the second-order operator $\cK_M=\cT_M\cT_M^\dagger$ by \eqref{eq:local-form}. Thus the need to measure derivative errors by a positive form leads to the second-order operator already used in our construction.

For a spinor $\Phi$ obeying $\mathsf L'$, \eqref{eq:approximation-norm} gives $M\|\Phi\|\leq\|\cT_M^\dagger\Phi\|$. Together with $\cT_\mu^\dagger=\cT_M^\dagger+\i(M-\mu)$, the triangle inequality therefore gives
\beq
\begin{aligned}
\|\cT_\mu^\dagger\Phi\|
\leq\|\cT_M^\dagger\Phi\|+|M-\mu|\,\|\Phi\|
\leq\left(1+\frac{|M-\mu|}{M}\right)q_M(\Phi,\Phi)^{1/2}.
\end{aligned}
\label{eq:mass-norm-bound}
\eeq
For $\Phi'$ also obeying $\mathsf L'$, the Schwarz inequality then gives
\beq
\big|\langle \cT_\mu^\dagger\Phi',\cT_M^\dagger\Phi\rangle\big|
\leq\left(1+\frac{|M-\mu|}{M}\right)
q_M(\Phi',\Phi')^{1/2}q_M(\Phi,\Phi)^{1/2}.
\label{eq:form-continuity-bound}
\eeq
Thus convergence in the norm defined by $q_M$ controls the form in \eqref{eq:local-form} for every fixed $\mu$.

We therefore use this norm to specify the approximation requirement. For finite-dimensional spaces $V_N$ of spinors obeying $\mathsf L'$, we require
\beq
\lim_{N\to\infty}\inf_{\Phi_N\in V_N}
q_M(\Phi-\Phi_N,\Phi-\Phi_N)=0
\label{eq:form-approximation}
\eeq
for every spinor $\Phi$ obeying $\mathsf L'$ with $q_M(\Phi,\Phi)<\infty$. The infimum is the smallest approximation error attainable within $V_N$, measured by the positive quadratic form.

To see why \eqref{eq:form-approximation} gives the regulated determinant of Section~\ref{sec:PV}, return to the physical field $\Psi=\cT_M^\dagger\Phi$. The error measured by $q_M^{1/2}$ is exactly the $L^2$ error in $\Psi$, and by \eqref{eq:approximation-norm} it also controls the error in $\Phi$. Completing the smooth spinors obeying $\mathsf L'$ in the $q_M^{1/2}$ norm gives the domain of definition of $\cT_M^\dagger$, denoted $\operatorname{Dom}(\cT_M^\dagger)$.\footnote{In Sobolev notation, $\operatorname{Dom}(\cT_M^\dagger)=\{\Phi\in H^1(Y)\mid P_+(\Phi|_W)=0\}$.} Since $\cT_M^\dagger\colon\operatorname{Dom}(\cT_M^\dagger)\to L^2(Y)$ is onto, \eqref{eq:form-approximation} says that the spaces $W_N=\cT_M^\dagger V_N$ approximate every $L^2$ spinor.

Writing $\Psi'=\cT_M^\dagger\Phi'$ for the transformed test spinor and recalling that $R=\cT_M^{-1}$, we have
\beq
\begin{aligned}
\langle\cT_\mu^\dagger\Phi',\cT_M^\dagger\Phi\rangle
&=\langle[1+\i(M-\mu)R^\dagger]\Psi',\Psi\rangle\\
&=\langle\Psi',[1-\i(M-\mu)R]\Psi\rangle .
\end{aligned}\label{eq:form-physical-ratio}
\eeq
In an orthonormal basis of $W_N$, the denominator ($\mu=M$) is the identity. If $P_N$ is the orthogonal projection onto $W_N$, the determinant ratio at finite $N$ is therefore $\det_{W_N}[1-\i(M-\mu)R_N]$, where $R_N=P_N R P_N$.

Removing the cutoff requires the spaces $W_N$ to approximate all of $L^2(Y)$. Given any $\Psi\in L^2(Y)$, choose $\Phi$ with $\cT_M^\dagger\Phi=\Psi$. Since $P_N\Psi$ is the best approximation to $\Psi$ in $W_N=\cT_M^\dagger V_N$, its squared error is
\beq
\begin{aligned}
\|\Psi-P_N\Psi\|_{L^2}^2
&=\inf_{\Phi_N\in V_N}\|\cT_M^\dagger(\Phi-\Phi_N)\|_{L^2}^2\\
&=\inf_{\Phi_N\in V_N}q_M(\Phi-\Phi_N,\Phi-\Phi_N).
\end{aligned}\label{eq:physical-projection-error}
\eeq
The right-hand side tends to zero by \eqref{eq:form-approximation}, so $P_N\Psi\to\Psi$ for every $L^2$ spinor. The Pauli--Villars regularization then proceeds as in Section~\ref{sec:PV}.

The basis functions must allow the boundary values left free by $\mathsf L'$. For example, on the interval above the transformed field $\Phi$ obeys $\Phi(0)=0$, while $\Phi(-\ell)$ is free. Requiring all basis functions to vanish at both ends would fail this requirement.

The Robin condition has a different role. It follows from varying the form in \eqref{eq:local-form} and need not be imposed on the basis functions. Moreover, for $\Phi$ obeying $\mathsf L'$, the formula \eqref{eq:green} gives $0\leq M\int_W|\Phi|^2\leq 2M\|\Phi\|\,\|\cD_Y\Phi\|\leq M^2\|\Phi\|^2+\|\cD_Y\Phi\|^2$. Thus the bulk terms alone define a norm equivalent to $q_M(\Phi,\Phi)^{1/2}$. For these reasons, the Robin condition plays no essential role in approximating fields in this norm. 

To evaluate the phase of the partition function numerically, we can therefore use a common finite-dimensional space $V_N$ for the numerator and denominator of \eqref{eq:ZN} without imposing the Robin condition. We choose the basis functions to obey $\mathsf L'$ and the spaces to satisfy the approximation criterion \eqref{eq:form-approximation} as $N$ increases. For each pair of basis functions $\Phi_j,\Phi_k$, the matrix entry is $\langle\cT_\mu^\dagger\Phi_j,\cT_M^\dagger\Phi_k\rangle$. For $\mu=M$, \eqref{eq:approximation-norm} gives $q_M(\Phi,\Phi)\geq M^2\|\Phi\|^2$ on $V_N$, so the denominator matrix is positive definite and the phase of the ratio comes entirely from the numerator.

\section{Lattice formulation}\label{sec:lattice}

We take the continuum formulation of Section~\ref{sec:continuum} as the starting point for lattice regularization. Its boundary conditions \eqref{eq:K-bc} depend only on the reference mass $M$, so the same field space serves the physical fermion and all its regulators. 

The leading $p^2$ term of $\cK_m$ makes its discretization natural. As seen in \eqref{eq:mp-intro}, this term is even in momentum and positive, whereas the Dirac term $\i\slashed p$ is odd. This even term also has the symmetries of a mass term. The lattice formulation discretizes $\cK_\mu$ and its boundary conditions, not the first-order operator $\cT_\mu$.

To make the discussion concrete, we restrict the continuum spacetime to the flat slab $Y=[-\ell,0]\times T^d$ and set $W_0=T^d$.

We take a hypercubic lattice $\Lambda=\{0,1,\ldots,n\}\times\Lambda_W$ with spacing $a$, approximating $Y$. Here $\Lambda_W$ is a periodic $d$-dimensional lattice discretizing $W_0$, and the layer $t \in \{0,1,\ldots,n\}$ sits at $\tau=(t-n)a$. The layers $t=n$ and $t=0$ are the two boundary components, at $\tau=0$ and $\tau=-\ell$ with $\ell=na$. A spin structure can be included in the links around the periodic directions. We write $\hat\nu$, $\nu=0,1,\ldots,d$, for the unit vectors, with $\hat0=\hat\tau$ along $+\tau$. We use $i=1,\ldots,d$ to label the directions along $W_0$ and write $\gamma^0=\gamma^\tau$. Unitary gauge links $U_\nu(x)$ live on the links of $\Lambda$ and commute with the gamma matrices.

\subsection{The bulk operator and Wilson's term}\label{sec:lat-bulk}

A spinor field $\phi(x)$, the lattice counterpart of $\Phi$, lives on the sites, where $\Phi$ is related to the original physical field $\Psi$ by $\Psi=\cT_M^\dagger\Phi=(\cD_Y-\i M)\Phi$, as in \eqref{eq:change}.

Away from the boundary layers, the lattice operator is
\beq
\cK^{\rm lat}_\mu=\Delta+\cR_a-\i(M-\mu)\cD_{\rm n}+\mu M. \label{eq:K-lat}
\eeq
This is a direct discretization of the continuum operator \eqref{eq:K}. We now explain the first three terms in \eqref{eq:K-lat} in order.

The first term, $\Delta$, discretizes the covariant Laplacian $-D^\nu D_\nu$. We use the covariant nearest-neighbor Laplacian with the nonnegative sign convention, defined by
\beq
\Delta\phi(x)=\frac1{a^2}\sum_\nu\big[2\phi(x)-U_\nu(x)\phi(x+a\hat\nu)-U_\nu(x-a\hat\nu)^\dagger\phi(x-a\hat\nu)\big].
\eeq

The second term, $\cR_a$, discretizes the spin coupling $\cR=-\sum_{\nu<\lambda}\gamma^{\nu\lambda}[D_\nu,D_\lambda]$. To represent these commutators on the lattice, we first define the centered covariant differences $\nabla_\nu$ in the bulk by
\beq
(\nabla_\nu\phi)(x)=\frac{U_\nu(x)\phi(x+a\hat\nu)-U_\nu(x-a\hat\nu)^\dagger\phi(x-a\hat\nu)}{2a}.
\eeq
The lattice spin coupling $\cR_a$ is then defined by
\beq
\cR_a\phi=-\sum_{\nu<\lambda}\gamma^{\nu\lambda}[\nabla_\nu,\nabla_\lambda]\phi
\qquad (1<t<n-1). \label{eq:R-lat-bulk}
\eeq
The layers $t=1,n-1$ adjacent to the boundary layers are excluded here, because the commutators on them involve the field on the boundary layers. They are treated in Section~\ref{sec:lat-boundary}. For free links, $[\nabla_\nu,\nabla_\lambda]=0$, so \eqref{eq:R-lat-bulk} vanishes. In the interior of a smooth background $[\nabla_\nu,\nabla_\lambda]\Phi=[D_\nu,D_\lambda]\Phi+O(a^2)$, so $\cR_a$ has the required continuum limit. Together, $\Delta+\cR_a$ represents $\cD_Y^2=-D^\nu D_\nu+\cR$.

The third term, $-\i(M-\mu)\cD_{\rm n}$, discretizes $-\i(M-\mu)\cD_Y$. Here $\cD_{\rm n}$ is the naive lattice Dirac operator, defined by
\beq
\cD_{\rm n}\phi(x)=\frac{\i}{2a}\sum_\nu\gamma^\nu\big[U_\nu(x)\phi(x+a\hat\nu)-U_\nu(x-a\hat\nu)^\dagger\phi(x-a\hat\nu)\big].
\eeq
Thus $\cD_{\rm n}=\i\sum_\nu\gamma^\nu \nabla_\nu$. Section~\ref{sec:lat-boundary} completes these definitions at the boundary.

Unlike its continuum counterpart \eqref{eq:K-def}, $\cK^{\rm lat}_\mu$ is not a product of two first-order lattice operators, since $\Delta$ differs from the square of $\cD_{\rm n}$. Properties that follow from that factorization in the continuum, such as the correspondence between zero modes of $\cK_m$ and of $\cT_m$, are therefore derived directly on the lattice below.

Wilson's term arises naturally. In flat continuum space without background fields, $-\i(M-\mu)\cD_Y=(M-\mu)\slashed\partial$, and $\cK_m$ for a plane wave of momentum $p$ is
\beq
\cK_m(p)=(M-m)\big[\i\slashed p+m(p)\big],\qquad m(p)=\frac{p^2+mM}{M-m}. \label{eq:mp}
\eeq
Up to the positive factor $M-m$, $\cK_m(p)$ is a Dirac operator whose mass depends on the momentum. At $p=0$ the mass is $mM/(M-m)$, which tends to $m$ as $M\to\infty$. On the mass shell $p^2=-m^2$ it equals $m$ exactly. It changes sign at $p^2=|m|M$, and at large momentum it grows like $p^2/(M-m)$.

On the lattice, $-\i(M-m)\cD_{\rm n}=(M-m)\gamma^\nu\nabla_\nu$. Hence, on every row except those of the boundary layers, $\cK^{\rm lat}_m$ can be written as
\beq
\begin{aligned}
&\cK^{\rm lat}_m=(M-m)\,D_{\rm Wilson}+\cR_a,\\
&D_{\rm Wilson}=\gamma^\nu\nabla_\nu+\frac{ra}{2}\Delta+m_0,\quad
r=\frac{2}{a(M-m)},\quad m_0=\frac{mM}{M-m}.
\end{aligned} \label{eq:wilson}
\eeq
This is an exact identity for arbitrary link variables. $D_{\rm Wilson}$ is Wilson's lattice Dirac operator~\cite{Wilson:1975id}, with Wilson parameter $r$ and bare mass $m_0$. Its Wilson term $(ra/2)\Delta$ represents the even term $p^2/(M-m)$ in \eqref{eq:mp}. The additional term $\cR_a$ represents the continuum spin coupling. For free fields it vanishes on the layers $1<t<n-1$, where \eqref{eq:R-lat-bulk} applies.

For free fields and a plane wave of momentum $p$,
\beq
\cK^{\rm lat}_m(p)=\hat p^2+mM+(M-m)\,\i\gamma^\nu\bar p_\nu,\qquad \bar p_\nu=\frac{\sin(ap_\nu)}{a},\quad \hat p^2=\sum_\nu\Big(\frac{2}{a}\sin\frac{ap_\nu}{2}\Big)^2, \label{eq:lat-mp}
\eeq
which is \eqref{eq:mp} with $p^2$ replaced by $\hat p^2$ in the mass and $p$ by $\bar p$ in the Dirac term. The odd term $\bar p$ vanishes at the $2^D$ momenta whose components are each $0$ or $\pi/a$, the corners of the Brillouin zone. The naive operator $\cD_{\rm n}$ alone would have a massless fermion at each of them: these are the doublers. Squaring $\cD_{\rm n}$ would not help, since $\bar p^2$ vanishes at every corner. The even term behaves differently: $\hat p^2$ vanishes only at $p=0$, and at a corner with $n_\pi$ components equal to $\pi/a$ it equals $4n_\pi/a^2$. Near such a corner $\cK^{\rm lat}_m(p)$ is $M-m$ times a Dirac operator with mass
\beq
m_{n_\pi}=\frac{4n_\pi/a^2+mM}{M-m}. \label{eq:corner-mass}
\eeq
At the origin this is $m_0<0$. At every other corner it is positive provided $a^2|m|M<4$. The sign of the mass therefore distinguishes the physical species from all the others, which is the lattice version of the statement that $m(p)$ is negative only at low momentum.

\subsection{Boundary conditions and the path integral}\label{sec:lat-boundary}

In the continuum, $\Phi$ obeys both the condition $\mathsf L'$ and the Robin condition in \eqref{eq:K-bc}. At $\tau=0$, these read $P_+\Phi=0$ and $(\partial_\tau+M)P_-\Phi=0$, respectively. We now explain how each condition is implemented on the lattice.

Recall that $t=0,\ldots,n$ labels the lattice layers, with $\tau=(t-n)a$. We first impose $\mathsf L'$ directly on the field values at the boundary layers,
\beq
P_+\phi(x)=0\ \text{on the layer}\ t=n,\qquad P_-\phi(x)=0\ \text{on the layer}\ t=0, \label{eq:lat-Lprime}
\eeq
where the opposite projectors at the two boundary layers reflect the outward normal $-\hat\tau$ at $t=0$.

We next implement the Robin condition of \eqref{eq:K-bc}, discretized by the difference between a boundary layer and the adjacent interior layer. Write $\phi_t$ for the field on the layer $t$ at a given position along $W_0$, and $U_\tau(t)$ for the normal link between the layers $t$ and $t+1$ at that position. On the top layer $t=n$, $(\partial_\tau+M)P_-\Phi=0$ becomes
\beq
\frac{P_-\phi_n-U_\tau(n-1)^\dagger P_-\phi_{n-1}}{a}
=-(1-\alpha)M\,P_-\phi_n-\alpha M\,U_\tau(n-1)^\dagger P_-\phi_{n-1}, \label{eq:lat-robin}
\eeq
where $\alpha$ is a real parameter. It specifies how the term $MP_-\Phi$ is divided between the layers $t=n$ and $t=n-1$. On the bottom layer $t=0$, the outward normal is $-\hat\tau$, and the same rule holds with $P_-$ replaced by $P_+$ and the layer $t=n-1$ by the layer $t=1$. The rule \eqref{eq:lat-robin} involves $M$ and $\alpha$ but not $\mu$. Like the continuum boundary conditions, it is the same for the physical field and for every regulator field.

Together, \eqref{eq:lat-Lprime} and \eqref{eq:lat-robin} determine the field on the boundary layers from the field on the adjacent interior layers. Solving \eqref{eq:lat-robin} for $P_-\phi_n$, which requires $1+(1-\alpha)aM\neq0$, and treating the layer $t=0$ in the same way, we obtain
\beq
\phi_n=\zeta_\alpha\,U_\tau(n-1)^\dagger P_-\phi_{n-1},\qquad
\phi_0=\zeta_\alpha\,U_\tau(0)P_+\phi_1,\qquad
\zeta_\alpha=\frac{1-\alpha aM}{1+(1-\alpha)aM}. \label{eq:lat-space}
\eeq
We denote by $\cV$ the space of lattice fields obeying \eqref{eq:lat-space}. A field in $\cV$ is determined by its values on the interior layers $0<t<n$, which are the independent variables. The conjugate field is $\overline\psi$, with $\overline\psi{}^\dagger$ in $\cV$, so that both fields have the same independent variables, just as $\Phi$ and $\overline\Psi{}^\dagger$ are expanded in the same functions $u_j$ in \eqref{eq:expansion}.

The parameter $\alpha$ enters only through $\zeta_\alpha$. For fixed $\alpha$ and $M$, $\zeta_\alpha=1-aM+O(a^2)$ as $a\to0$, so the choice of $\alpha$ does not affect the continuum limit of \eqref{eq:lat-space}. The reason that we introduced $\alpha$ as a free parameter rather than a fixed number (such as $1/2$) is as follows. If $M$ is of order $a^{-1}$, the choice $\alpha aM=1$ gives $\zeta_\alpha=0$. The field then vanishes on the boundary layers $t=0,n$, and the lattice of independent variables simply ends, as in the boundary-fermion construction of~\cite{Shamir:1993zy}. We compare the two formulations in Section~\ref{sec:related-work}.

The operator \eqref{eq:K-lat} acts on the interior layers $0<t<n$. On the layers $t=1$ and $t=n-1$, the terms $\Delta$ and $-\i(M-\mu)\cD_{\rm n}$ involve the field on the adjacent boundary layer, which is given by \eqref{eq:lat-space}.

It remains to specify the spin coupling $\cR_a$ on the layers $t=1$ and $t=n-1$, where the normal difference $\nabla_\tau$ involves the boundary values \eqref{eq:lat-space}. In the continuum, $\cR$ is tied to the kinetic terms by the identity
\beq
\|\cD_Y\Phi\|^2=\sum_\nu\|D_\nu\Phi\|^2+\langle\Phi,\cR\,\Phi\rangle\qquad(\Phi\ \text{obeys}\ \mathsf L'), \label{eq:continuum-spin-identity}
\eeq
which was used to derive \eqref{eq:local-form}. We define $\cR_a$ by requiring that the lattice version of this identity holds,
\beq
\left\|\sum_\nu\gamma^\nu \nabla_\nu\phi\right\|^2
=\sum_\nu\|\nabla_\nu\phi\|^2+\langle\phi,\cR_a\phi\rangle,
\qquad \phi\in\cV. \label{eq:lat-spin-identity}
\eeq
Here $\langle\phi',\phi\rangle=\sum_{x:\,0<t<n}(\phi'(x))^\dagger\phi(x)$ is the inner product on $\cV$, with the sum running over the interior layers, and $\|\phi\|^2=\langle\phi,\phi\rangle$. The differences $\nabla_\nu\phi$ are evaluated on the interior layers, and their norms include all spinor components. Since $\cV$ is a complex vector space, \eqref{eq:lat-spin-identity} for every $\phi\in\cV$ determines the operator $\cR_a$ uniquely, and $\cR_a$ is hermitian because the other two terms are real. The positivity proof in Section~\ref{sec:lat-positive} relies on this definition.

To write $\cR_a$ explicitly, recall that the continuum gauge field obeys $A_\tau=0$ and $\partial_\tau A_i=0$ near the boundary, as assumed in Section~\ref{sec:setup}. We impose the corresponding conditions on the lattice links,
\beq
U_\tau(x)=1,\qquad U_i(x+a\hat\tau)=U_i(x),\qquad i=1,\ldots,d, \label{eq:lat-gauge-collar}
\eeq
in a neighborhood of each boundary wide enough to contain the links used by $\nabla_\nu$ on the layers $t=1,n-1$.

We evaluate the definition \eqref{eq:lat-spin-identity} by transferring $\nabla_\nu$ from $\phi$ to a test spinor $\phi'$ in the sum over the interior layers, just as in integration by parts. This discrete analog is called summation by parts. For arbitrary lattice spinors $\phi$ and $\phi'$ on $\Lambda$, without imposing boundary conditions, the formulas under \eqref{eq:lat-gauge-collar} are
\beq
\begin{aligned}
\sum_{x:\,0<t<n}(\phi'(x))^\dagger(\nabla_i\phi)(x)
&=-\sum_{x:\,0<t<n}(\nabla_i\phi')(x)^\dagger\phi(x),\\
\sum_{0<t<n}(\phi'_t)^\dagger(\nabla_\tau\phi)_t
&=-\sum_{0<t<n}(\nabla_\tau\phi')_t^\dagger\phi_t\\
&\quad+\frac1{2a}\Big[(\phi'_{n-1})^\dagger\phi_n+(\phi'_n)^\dagger\phi_{n-1}\Big]
-\frac1{2a}\Big[(\phi'_1)^\dagger\phi_0+(\phi'_0)^\dagger\phi_1\Big],
\end{aligned} \label{eq:lat-SBP}
\eeq
where the second formula holds at each position along $W_0$. In the normal direction, the centered differences leave boundary terms that pair each boundary layer with the adjacent interior layer. They are the lattice form of the boundary terms obtained by integration by parts in the continuum.

Expanding the left-hand side of \eqref{eq:lat-spin-identity} and applying \eqref{eq:lat-SBP} gives the commutators of the differences $\nabla_\nu$ and a boundary term. On the layers $1<t<n-1$ only the commutators appear, and \eqref{eq:lat-spin-identity} reproduces the formula \eqref{eq:R-lat-bulk}. On the layers $t=1,n-1$, the differences $\nabla_i$ along $W_0$ commute with the normal difference $\nabla_\tau$ under \eqref{eq:lat-gauge-collar}, so $[\nabla_\tau,\nabla_i]=0$. The boundary term at $t=n$ is $\frac1{2a}\sum_i[(\phi_{n-1})^\dagger\gamma^\tau\gamma^i\nabla_i\phi_n+(\phi_n)^\dagger\gamma^\tau\gamma^i\nabla_i\phi_{n-1}]$, summed over the positions along $W_0$, and the boundary term at $t=0$ is the same expression for the layers $t=0,1$ with the opposite sign. Evaluating them with \eqref{eq:lat-space} gives
\beq
\cR_a\phi=-\sum_{1\leq i<j\leq d}\gamma^{ij}[\nabla_i,\nabla_j]\phi
\pm\frac{\zeta_\alpha}{2a}\sum_{i=1}^d\gamma^\tau\gamma^i\nabla_i\phi, \label{eq:R-lat-boundary}
\eeq
with the upper sign on the layer $t=n-1$ and the lower sign on the layer $t=1$.
The first term is gauge covariant and uses at most two links along $W_0$ within the same layer. The second term is the lattice form of the boundary term $\int_W\Phi^\dagger\gamma^\tau\gamma^iD_i\Phi$ that arises in the integration by parts leading to \eqref{eq:continuum-spin-identity}. In the continuum this boundary term vanishes, because $D_i$ preserves the chirality of $\Phi$ on $W$ whereas $\gamma^\tau\gamma^i$ reverses it. On the lattice it pairs a boundary layer with the adjacent interior layer, where both chiralities are present, and it does not vanish. It is present even for free links, and it is absent for $\zeta_\alpha=0$. For smooth fields $\Phi'$ and $\Phi$ obeying $\mathsf L'$, the component $P_+\Phi$ on the layer $t=n-1$ and the component $P_-\Phi$ on the layer $t=1$ are of order $a$, and likewise for $\Phi'$, so the second term contributes to $a^D\langle\Phi',\cR_a\Phi\rangle$ only at order $a$. Together with \eqref{eq:R-lat-bulk}, \eqref{eq:R-lat-boundary} gives $\cR_a$ on all interior layers.

With the boundary values \eqref{eq:lat-space} and the spin coupling $\cR_a$, \eqref{eq:K-lat} on the interior layers $0<t<n$ defines a linear map of $\cV$ to itself, and we denote its matrix on the independent variables by $\cK^{\rm lat}_\mu$ as well.

The lattice partition function is
\beq
Z^{\rm lat}=\frac{\det\cK^{\rm lat}_m}{\det\cK^{\rm lat}_M}. \label{eq:Zlat}
\eeq
The Grassmann integral over $\phi$ and $\overline\psi$ with action $\i\sum_x\overline\psi(x)(\cK^{\rm lat}_m\phi)(x)$, where $x$ runs over the interior layers $0<t<n$, gives $\det\cK^{\rm lat}_m$ up to a mass-independent normalization. Multiplying by the normalized commuting Gaussian at $\mu=M$, which supplies $1/\det\cK^{\rm lat}_M$, gives the ratio in \eqref{eq:Zlat}. The same space $\cV$ serves both, as $U_N$ did in \eqref{eq:ZN}. On a finite lattice the integrals are finite, and no further Pauli--Villars fields are needed for finiteness.

The mass $M$ can be used in two ways. If $M$ and $\alpha$ are held fixed as $a\to0$, every term of the lattice operator approximates its continuum counterpart in a smooth gauge background, with the boundary conditions of Section~\ref{sec:continuum}.
If $M$ is of the order of the cutoff, the regulator instead lives at the lattice scale. The choice $\alpha aM=1$ is of this second type.

In both prescriptions, the spin coupling in $\cK^{\rm lat}_m/(M-m)$ is $\cR_a/(M-m)$. Its commutator terms are suppressed by the regulator mass and vanish in a fixed smooth background when the regulator is removed, $M\to\infty$, after or together with $a\to0$. The second term of \eqref{eq:R-lat-boundary} is confined to the layers $t=1,n-1$. The effects of $\cR_a$ on the low-energy path integral are therefore expected to vanish up to local bulk and boundary counterterms.

\subsection{Positivity of the denominator}\label{sec:lat-positive}

We now show that the denominator $\cK^{\rm lat}_M$ is hermitian and positive. These properties imply that the denominator contributes no phase, so that the phase of $Z^{\rm lat}$ is that of $\det\cK^{\rm lat}_m$.
In the continuum, positivity \eqref{eq:KM-positive} follows from writing $\cK_M=\cT_M\cT_M^\dagger$. On the lattice we show it by writing $\phi^\dagger\cK^{\rm lat}_M\phi$ as an explicit sum of nonnegative terms.

For the proof, we use the inner product $\langle\phi',\phi\rangle$ and the norm $\|\phi\|$ on $\cV$ introduced below \eqref{eq:lat-spin-identity}, so that $\phi^\dagger\cK^{\rm lat}_M\phi=\langle\phi,\cK^{\rm lat}_M\phi\rangle$. We define the nonnegative quadratic form
\beq
E(\phi)=\sum_{(x,\nu)}
\frac{|U_\nu(x)\phi(x+a\hat\nu)-\phi(x)|^2}{a^2}, \label{eq:link-energy}
\eeq
where the sum runs over the links along $W_0$ within the interior layers $0<t<n$ and over all normal links of $\Lambda$. The normal links include those ending on the boundary layers $t=0,n$, where $\phi$ is given by \eqref{eq:lat-space}. Summing $\Delta$ by parts and using \eqref{eq:lat-space} gives
\beq
\phi^\dagger\cK^{\rm lat}_M\phi={}&E(\phi)+\langle\phi,\cR_a\phi\rangle
+M^2\|\phi\|^2\nonumber\\
&+\frac{\zeta_\alpha(1-\zeta_\alpha)}{a^2}\left[\sum_{x:\,t=n-1}|P_-\phi(x)|^2+\sum_{x:\,t=1}|P_+\phi(x)|^2\right]. \label{eq:lat-reference-form}
\eeq
The term $-\i(M-\mu)\cD_{\rm n}$ is absent because $\mu=M$. The last term is the boundary term of the summation by parts, evaluated with \eqref{eq:lat-space}. 

$E(\phi)$ is at least $\sum_\nu\|\nabla_\nu\phi\|^2$. To prove this bound, define the forward covariant difference by
\beq
(\nabla^+_\nu\phi)(x)=\frac{U_\nu(x)\phi(x+a\hat\nu)-\phi(x)}{a}. \label{eq:lat-forward-difference}
\eeq
At an interior site $x$, the centered difference can be written as
\beq
(\nabla_\nu\phi)(x)=\frac12\Big[(\nabla^+_\nu\phi)(x)
+U_\nu(x-a\hat\nu)^\dagger(\nabla^+_\nu\phi)(x-a\hat\nu)\Big], \label{eq:lat-link-average}
\eeq
and
\beq
\begin{aligned}
&\frac12\Big[|(\nabla^+_\nu\phi)(x)|^2+|(\nabla^+_\nu\phi)(x-a\hat\nu)|^2\Big]
-|(\nabla_\nu\phi)(x)|^2\\
&\qquad=\frac14\Big|(\nabla^+_\nu\phi)(x)
-U_\nu(x-a\hat\nu)^\dagger(\nabla^+_\nu\phi)(x-a\hat\nu)\Big|^2.
\end{aligned} \label{eq:lat-link-variance}
\eeq
Adding over the interior sites gives exactly
\beq
E(\phi)-\sum_\nu\|\nabla_\nu\phi\|^2
=\frac{a^2}{4}\sum_\nu\|\Delta_\nu\phi\|^2
+\frac12\sum_{x:\,t=0,n-1}|(\nabla^+_\tau\phi)(x)|^2. \label{eq:lat-remainder}
\eeq
Here $\Delta_\nu$ is the contribution to $\Delta$ from the direction $\nu$ alone, evaluated on the interior layers. The last term arises because a normal link ending on a boundary layer has only one end on an interior layer. Such a link is counted once in $E(\phi)$ but only with weight $1/2$ when \eqref{eq:lat-link-variance} is added over the interior sites. By \eqref{eq:lat-space}, a link between the layers $t=n-1$ and $t=n$ has $a^2|(\nabla^+_\tau\phi)(x)|^2=|P_+\phi_{n-1}|^2+(1-\zeta_\alpha)^2|P_-\phi_{n-1}|^2$, and a link between the layers $t=0$ and $t=1$ has $a^2|(\nabla^+_\tau\phi)(x)|^2=|P_-\phi_1|^2+(1-\zeta_\alpha)^2|P_+\phi_1|^2$.

Combining \eqref{eq:lat-spin-identity}, \eqref{eq:lat-reference-form} and \eqref{eq:lat-remainder}, we obtain
\beq
\phi^\dagger\cK^{\rm lat}_M\phi={}&\left\|\sum_\nu\gamma^\nu \nabla_\nu\phi\right\|^2
+\frac{a^2}{4}\sum_\nu\|\Delta_\nu\phi\|^2+M^2\|\phi\|^2\nonumber\\
&+\frac{1}{2a^2}\sum_{x:\,t=n-1}\Big[|P_+\phi(x)|^2+(1-\zeta_\alpha^2)|P_-\phi(x)|^2\Big]\nonumber\\
&+\frac{1}{2a^2}\sum_{x:\,t=1}\Big[|P_-\phi(x)|^2+(1-\zeta_\alpha^2)|P_+\phi(x)|^2\Big]. \label{eq:lat-positive}
\eeq
The right-hand side is real for every $\phi\in\cV$, so $\cK^{\rm lat}_M$ is hermitian. Every term is nonnegative when $|\zeta_\alpha|\leq1$, and then $\phi^\dagger\cK^{\rm lat}_M\phi\geq M^2\|\phi\|^2$. We assume $|\zeta_\alpha|\leq1$ in what follows.
Thus $\cK^{\rm lat}_M$ is hermitian and strictly positive for every $M>0$ and all unitary links.\footnote{The proof does not use \eqref{eq:lat-gauge-collar}. For links that do not satisfy it, \eqref{eq:lat-spin-identity} adds the commutators $-\sum_i\gamma^\tau\gamma^i[\nabla_\tau,\nabla_i]\phi$ to \eqref{eq:R-lat-boundary}.}
The determinant $\det\cK^{\rm lat}_M$ has no phase. Equation \eqref{eq:lat-positive} is the counterpart of \eqref{eq:KM-positive}. 

To interpret the boundary terms in \eqref{eq:lat-positive}, restore the volume factor $a^D$ and write each boundary sum with the measure $a^{D-1}$. The coefficient of $|P_+\phi_{n-1}|^2$ at $t=n-1$ and of $|P_-\phi_1|^2$ at $t=1$ is then $1/(2a)$. Since this coefficient diverges as $a\to0$, a finite continuum quadratic form requires $P_+\Phi=0$ at $\tau=0$ and $P_-\Phi=0$ at $\tau=-\ell$. These are precisely the boundary conditions $\mathsf L'$.
For the remaining components, $P_-\phi_{n-1}$ and $P_+\phi_1$, the coefficient is $(1-\zeta_\alpha^2)/(2a)$. For fixed $\alpha$ and $M$, this is $M+O(a)$, so these terms give $M\int_W|\Phi|^2$ in \eqref{eq:KM-positive}. As discussed at the end of Section~\ref{sec:sine-cutoff}, the bulk terms alone define an equivalent norm in the continuum. On the lattice, the size of this boundary term is likewise unimportant as long as the term is nonnegative. In particular, the cutoff-scale choice $\alpha aM=1$ gives $\zeta_\alpha=0$ and a positive coefficient $1/(2a)$.

\subsection{Local counterterms in the bulk}\label{sec:lat-counterterms}

The partition function depends on the bulk of $Y$ through local terms such as the vacuum energy and the gauge kinetic term. These local terms must be canceled by local counterterms for the partition function to depend only on the boundary data in the large-mass limit \eqref{eq:chiral-limit}. This section examines which local bulk terms the lattice partition function \eqref{eq:Zlat} can contain. Since these terms are the same with and without a boundary, we consider a manifold or a lattice without boundary. Terms localized on the boundary are not considered. The discussion is at the level of power counting, and is not a mathematical proof.

We begin with the continuum. Let $Y$ be closed. Then $\cD_Y$ is self-adjoint with real eigenvalues $\lambda$, the positive operator $\cK_M$ equals $\cD_Y^2+M^2$, and the space $U_N$ of Section~\ref{sec:measure} is spanned by the eigenvectors of $\cD_Y$ with $|\lambda|$ below a bound. In this basis $\cK_\mu=(\cD_Y+\i\mu)(\cD_Y-\i M)$ is diagonal, and the ratio \eqref{eq:ZN} becomes
\beq
Z_N(m;M)=\prod_{\lambda}\frac{\lambda+\i m}{\lambda+\i M}, \label{eq:closed-product}
\eeq
where the product runs over the retained eigenvalues. For $m=-M$ every factor has modulus one, so the bulk partition function is a pure phase. The vacuum energy, the gauge kinetic term and every other real local term then cancel between the physical fermion and the Pauli--Villars field, and no counterterm is needed for the real part of $\log Z_N$. The phase is $-2\sum_\lambda\arctan(M/\lambda)$, which is $-\pi$ times the spectral asymmetry $\sum_\lambda\sign(\lambda)$ with the eigenvalues above the scale $M$ smoothly cut off and with $\sign(0)=+1$. This is how $\exp(-2\pi\i\eta_D)$ in \eqref{eq:WY-intro} arises in the large-mass limit.

On the lattice, we correspondingly consider a lattice that is periodic in all $D$ directions, so that every row of the operator is given by \eqref{eq:K-lat} and $Z^{\rm lat}=\det\cK^{\rm lat}_m/\det\cK^{\rm lat}_M$. The continuum property $|Z_N|=1$ at $m=-M$ is not exact on such a lattice. For free links, \eqref{eq:lat-mp} with $m$ replaced by $\mu$ gives
\beq
\cK^{\rm lat}_\mu(p)^\dagger\,\cK^{\rm lat}_\mu(p)
=(\hat p^2+\mu^2)(\hat p^2+M^2)-\frac{a^2(M-\mu)^2}{4}\sum_\nu\hat p_\nu^4,
\qquad \hat p_\nu=\frac2a\sin\frac{ap_\nu}{2}. \label{eq:lat-modulus}
\eeq
The first term is the lattice form of the continuum expression $(p^2+\mu^2)(p^2+M^2)$, which depends on $\mu$ only through $\mu^2$. The second term arises because $\hat p^2$ exceeds $\bar p^2$ by $\tfrac{a^2}{4}\sum_\nu\hat p_\nu^4$. This is the term that removes the doublers. It does not vanish at $\mu=-M$, so $|Z^{\rm lat}|\neq1$ at $m=-M$ already for free links.

The local terms are nevertheless strongly constrained. Three properties hold exactly on a lattice without boundary, for arbitrary link variables.
\begin{itemize}
\item At $m=M$ the numerator and the denominator coincide, so $Z^{\rm lat}=1$.
\item The operators $\Delta$, $\cR_a$ and $\cD_{\rm n}$ are hermitian. The adjoint of $\cK^{\rm lat}_\mu$ is therefore obtained from \eqref{eq:K-lat} by the formal replacement $(\mu,M)\to(-\mu,-M)$, and $Z^{\rm lat}$ goes to its complex conjugate under $(m,M)\to(-m,-M)$.
\item Let $U^R$ be the link configuration obtained from $U$ by the reflection of one coordinate $x^\nu$. The reflection of the sites, combined with the multiplication of the field by $\gamma^\nu$, maps $\cK^{\rm lat}_\mu$ in the background $U$ to the operator \eqref{eq:K-lat} with $(\mu,M)$ replaced by $(-\mu,-M)$ in the background $U^R$. Together with the previous property, $Z^{\rm lat}[U^R]=\overline{Z^{\rm lat}[U]}$.
\end{itemize}
The continuum expression \eqref{eq:closed-product} has the same three properties. Under either the reflection $U\to U^R$ or the mass sign reversal $(m,M)\to(-m,-M)$, $\log|Z^{\rm lat}|$ is unchanged and the phase of $Z^{\rm lat}$ changes sign. At $m=M$, we have $Z^{\rm lat}=1$, so $\log|Z^{\rm lat}|=0$ and the phase vanishes.

The reflection property is not limited to a lattice without boundary. For $\alpha aM=1$ it also holds on the slab $\Lambda$ with its two boundary layers $t=0,n$. In this case $\zeta_\alpha=0$ and the hopping to the boundary layers is omitted, so $\Delta$, $\cR_a$ and $\cD_{\rm n}$ remain hermitian and the derivation above applies, both for the reflection $t\to n-t$ and for the reflection of a coordinate along $W_0$. Thus $Z^{\rm lat}[U^R]=\overline{Z^{\rm lat}[U]}$ is exact on $\Lambda$, and in particular $Z^{\rm lat}$ is real for links that are independent of $t$ with $U_\tau=1$. For $\zeta_\alpha\neq0$ a reflection does not preserve the boundary values \eqref{eq:lat-space}, and this relation holds only up to lattice artifacts localized at the boundaries.

A counterterm is needed for a local term of $\log Z^{\rm lat}$ whose coefficient does not vanish when the lattice spacing is removed. Such a term has the form $c_\cO\int\cO$, where $\cO$ is a gauge-invariant local operator built from the gauge field, and the coefficient $c_\cO$ has mass dimension $D-\dim\cO$. By the three properties, a term in the real part of $\log Z^{\rm lat}$ has an operator $\cO$ that is even under reflection and a coefficient that is even under $(m,M)\to(-m,-M)$. A term in the imaginary part has an odd operator and an odd coefficient. All coefficients vanish at $m=M$.

To estimate the coefficients, suppose that $m$ and $M$ are held fixed as $a\to0$. The lattice and the continuum expressions then differ through the momenta of order $1/a$, where the masses are small perturbations. The contribution of these momenta to $c_\cO$ is a polynomial in $m$ and $M$, in which each power of a mass is accompanied by a power of $a$. A term of second degree in the masses is thus of order $a^{\dim\cO+2-D}$, and a term of first degree is of order $a^{\dim\cO+1-D}$. This is ordinary power counting, in which each power of a mass lowers the degree of divergence by one.

Consider first the real part. The polynomial is even and vanishes at $m=M$, so its lowest terms are of second degree, with the two structures $m^2-M^2$ and $(M-m)^2$. The first is the lattice form of the continuum divergence and vanishes at $m=-M$. The second is the lattice artifact seen in \eqref{eq:lat-modulus} and does not vanish at $m=-M$. These terms survive as $a\to0$ only for $\dim\cO\leq D-2$, and terms of higher degree only for still smaller $\dim\cO$. For $D\leq5$ the only such operator is the identity, since the gauge kinetic term has dimension four. The momenta of order $1/a$ then contribute only a vacuum energy, of order $a^{2-D}$ up to logarithms and independent of the gauge field. The remaining real local terms are those of the continuum theory with the same masses, which vanish at $m=-M$ by \eqref{eq:closed-product}. Thus, for $|m|=M$ and $D\leq5$, no gauge kinetic term is induced in the bulk as $a\to0$, and the only counterterm needed for the real part is a constant vacuum energy. For $|m|\neq M$ the local terms of the continuum theory, such as the gauge kinetic term for $D\geq4$, must be canceled as well.

Consider next the imaginary part. The polynomial is odd under $(m,M)\to(-m,-M)$, so its lowest terms are of first degree. Since the polynomial also vanishes at $m=M$, these terms are proportional to $M-m$. They survive as $a\to0$ only for $\dim\cO\leq D-1$. There is no gauge-invariant local operator of dimension less than $D$ that is odd under reflection. The operators of lowest dimension, such as the Chern--Simons term for odd $D$, have dimension $D$. Hence the momenta of order $1/a$ do not contribute to the phase as $a\to0$, and no counterterm is needed for the imaginary part. The phase is that of the continuum theory with the same masses, which becomes $-2\pi\eta_D$ in the large-mass limit.

Operators of dimension greater than $D$ also appear, with coefficients suppressed by powers of $a$ and of the inverse masses. At finite lattice spacing and finite masses they make the phase of $Z^{\rm lat}$ differ from $-2\pi\eta_D$. The difference disappears when the limit $a\to0$ is followed by the large-mass limit.

If instead $aM$ is held fixed, the masses are not small at momenta of order $1/a$, and the coefficients are functions of $am$ and $aM$ that are not polynomials. The three exact properties still hold. Dimensional analysis then allows in the real part a vacuum energy and, for $D\geq4$, a gauge kinetic term, even at $m=-M$, and both require counterterms in general. In the imaginary part there is still no operator of dimension less than $D$.

\paragraph{Exact cancellation with a square root.} The real local terms of the bulk can be canceled exactly, at finite lattice spacing, if the positive denominator of \eqref{eq:Zlat} is replaced by an expression with a square root. The new denominator must satisfy the following requirements.
\begin{itemize}
\item It is positive.
\item It has no boundary mode.
\item On a lattice without boundary, it equals $|\det\cK^{\rm lat}_m|$.
\end{itemize}
The last requirement makes the partition function have modulus one on a lattice without boundary. The first two ensure that, on $\Lambda$ with its two boundary layers $t=0,n$, the phase and boundary fermions remain those of $\det\cK^{\rm lat}_m$, as for \eqref{eq:Zlat}.

We construct such a denominator using a doubled lattice. Let $\widehat\Lambda$ be the lattice without boundary obtained by joining $\Lambda$ to its mirror image under the reflection of $t$ along the layers $t=0$ and $t=n$, with the reflected links $U^R$ on the mirror image, and let $\widehat\cK^{\rm lat}_m$ be the operator \eqref{eq:K-lat} on $\widehat\Lambda$. The determinant $\det\cK^{\rm lat}_m$ is complex conjugated under reflection, by the reflection properties of $\cK^{\rm lat}_\mu$ stated above. Since the link configuration of $\widehat\Lambda$ is invariant under reflection, $\det\widehat\cK^{\rm lat}_m$ is real. Since the real local terms are even under reflection, the two halves of $\widehat\Lambda$ contribute them equally. The denominator $|\det\widehat\cK^{\rm lat}_m|^{1/2}$ therefore cancels the real local bulk terms of $\det\cK^{\rm lat}_m$ on $\Lambda$. A subtraction of this type was used for domain-wall fermions in~\cite{Kikukawa:2001mw}.

The role of the doubled lattice can also be understood from the corresponding continuum construction. For a general compact manifold $Y$ with boundary, the double $Y\cup_{\partial Y}\overline Y$ is obtained by gluing $Y$ to its reflected copy $\overline Y$, with the corresponding reflected spin and background structures. This construction plays an important role in reflection positivity, which requires $Z(Y\cup_{\partial Y}\overline Y)\geq0$ because this partition function is the squared norm of the state prepared on $Y$~\cite{Freed:2016rqq,Yonekura:2018ufj}. Taking its square root therefore gives the norm of that state. See also~\cite{Kontsevich:2021dmb} for reflection positivity on curved backgrounds beyond topological field theory.

This relation between doubling and taking a norm motivates the subtraction above. Given an invertible field theory with partition function $Z$, its absolute value $|Z|$ also defines an invertible field theory which is topologically trivial. This theory is described by a real local action $S_{\rm loc}=-\log|Z|$. Dividing by $|Z|$ therefore amounts to adding the counterterm $-S_{\rm loc}$ and leaves the phase $Z/|Z|$. Since the theory defined by $|Z|$ is topologically trivial, it does not require boundary modes, so this subtraction is compatible with preserving the physical boundary fermions.

\subsection{One chiral fermion on each boundary}\label{sec:lat-chiral}

We now isolate one boundary by taking the half-space limit ($-\ell \to -\infty$) and discuss whether it carries exactly one chiral fermion.
Here one chiral fermion means one massless species localized near $t=n$. Let $k$ denote the momentum along $W_0$. At $k=0$, $\cK^{\rm lat}_m$ must have one normalizable zero-mode profile with $\gamma^\tau=+1$. A corresponding normalizable zero-mode profile of $(\cK^{\rm lat}_m)^\dagger$ with $\gamma^\tau=-1$ supplies the conjugate field.
At small $k$, the kinetic operator between these modes must have a nonzero term proportional to $\i P_-\gamma^iP_+k_i$. There must be no additional zero at $k\ne0$ anywhere in the boundary Brillouin zone.

Consider free fields below the top boundary layer $t=n$. Let $j=t-n=0,-1,-2,\ldots$ label the layers, so $\tau=ja$, and write $\phi^+_j=P_+\phi$ and $\phi^-_j=P_-\phi$ for the two chirality components of $\phi$ on the layer $j$. The boundary values \eqref{eq:lat-space} are $\phi^+_0=0$ and $\phi^-_0=\zeta_\alpha\phi^-_{-1}$, and the independent variables are $\phi_j$ with $j\leq-1$. The part of $\cD_{\rm n}$ acting along $W_0$ is $-a^{-1}\sum_i\gamma^i\sin(ak_i)$, which vanishes at the $2^d$ corners of the boundary Brillouin zone, where each $k_i$ is $0$ or $\pi/a$. The second term of \eqref{eq:R-lat-boundary}, which is present on the layer $j=-1$ even for free fields, is proportional to $\sum_i\gamma^\tau\gamma^i\sin(ak_i)$ and vanishes at these corners as well. There $\phi^+$ and $\phi^-$ decouple.

In the bulk, set $\beta=a(M-m)/2$. The differences along $\tau$ in $-\i(M-m)\cD_{\rm n}$ contribute $\beta\gamma^\tau(\phi_{j+1}-\phi_{j-1})$ to $a^2\cK^{\rm lat}_m\phi$. The diagonal entry of $a^2\cK^{\rm lat}_m$ is $b=2+a^2(\hat k^2+mM)$, where $\hat k^2=\sum_i[2\sin(ak_i/2)/a]^2$. At a corner of the boundary Brillouin zone ($k_i\in\{0,\pi/a\}$ for all $i$), $a^2\cK^{\rm lat}_m\phi=0$ gives, for $\phi^+$,
\beq
b\,\phi^+_j-(1+\beta)\phi^+_{j-1}-(1-\beta)\phi^+_{j+1}=0. \label{eq:positive-chirality-recurrence}
\eeq
The solutions are $\phi^+_j=y^{-j}$ with
\beq
(1+\beta)y^2-by+1-\beta=0. \label{eq:roots}
\eeq
The mode decays as $j\to-\infty$ when $|y|<1$.
The equation for $\phi^-$ has the opposite sign of $\gamma^\tau$, so the coefficients $1+\beta$ and $1-\beta$ are exchanged, and its roots are the reciprocals of those of \eqref{eq:roots}.

Before considering individual corners, we determine when the decaying bulk solutions can satisfy the boundary conditions. For the moment, we assume $0<\beta<1$, or $0<a(M-m)<2$, so that both hopping coefficients $1\pm\beta$ are positive. We treat the endpoint $\beta=1$ separately below. A nonzero $\phi^+$ requires two decaying solutions, since the boundary condition $\phi^+_0=0$ removes a single decaying solution. If both roots $y_1,y_2$ of \eqref{eq:roots} lie between $0$ and $1$, this condition selects one combination, $\phi^+_j\propto y_1^{-j}-y_2^{-j}$.

To locate the roots, use $\beta^2+a^2mM=a^2(M+m)^2/4\geq0$, which gives $a^2mM\geq-\beta^2$. With $\beta<1$, it follows that $b\geq2-\beta^2>0$ and the discriminant obeys $b^2-4(1-\beta^2)\geq\beta^4>0$. The two roots are therefore distinct and positive. Their product is $y_1y_2=(1-\beta)/(1+\beta)$, and $\beta>0$ makes it less than $1$, so they cannot both exceed $1$. Since the polynomial in \eqref{eq:roots} equals $2-b$ at $y=1$, both roots lie below $1$ precisely when $b<2$. For $b>2$, one root lies below $1$ and the other above $1$. At $b=2$, one root equals $1$ and does not decay.

For $\phi^-$, the roots are $1/y_1$ and $1/y_2$, so their product is greater than $1$ and at most one solution decays. If such a solution exists, it has the form $\phi^-_j={y'}^{-j}\phi^-_0$ with $0<y'<1$. The boundary value $\phi^-_0=\zeta_\alpha\phi^-_{-1}$ in \eqref{eq:lat-space} then requires $\zeta_\alpha y'=1$ for a nonzero solution. This is impossible for $|\zeta_\alpha|\leq1$, the condition assumed in Section~\ref{sec:lat-positive}. Thus $\phi^-$ cannot support a boundary zero mode.

At the endpoint $\beta=1$, or $a(M-m)=2$, the hopping coefficient $1-\beta$ vanishes and the $\phi^+$ recurrence no longer involves $\phi^+_{j+1}$. With $\phi^+_0=0$, its solution is $\phi^+_j=(b/2)^{-j-1}\phi^+_{-1}$ for $j\leq-1$, which decays precisely when $b<2$. The $\phi^-$ root is $2/b$, and whenever it decays it is again excluded by $|\zeta_\alpha|\leq1$. Hence, throughout $0<\beta\leq1$, a corner supports a boundary zero mode if and only if
\beq
b<2\qquad\Longleftrightarrow\qquad \hat k^2+mM<0. \label{eq:corner-zero-condition}
\eeq
When it exists, the zero mode has one independent profile and $\gamma^\tau=+1$.

At $k=0$, \eqref{eq:corner-zero-condition} reduces to $mM<0$. Since $M>0$, a boundary zero mode therefore requires $m<0$. With this choice, $b=2+a^2mM<2$ and there is exactly one zero mode with positive chirality. For $0<\beta<1$, the boundary condition $\phi^+_0=0$ selects the profile
\beq
\phi^+_j\propto y_1^{-j}-y_2^{-j}.
\eeq
At $\beta=1$, the profile is $\phi^+_j=(b/2)^{-j-1}\phi^+_{-1}$ for $j\leq-1$.

To compare this lattice mode with the physical boundary mode in Section~\ref{sec:setup}, consider the positive-chirality solutions of the continuum equation $\cK_m\Phi=0$ at $k=0$ on the half-line $\tau\leq0$. For $|m|\ne M$, the two decaying profiles are $e^{|m|\tau}$ and $e^{M\tau}$, the counterparts of the two lattice solutions for $0<\beta<1$. As on the lattice, the boundary condition $\Phi(0)=0$ selects their difference.\footnote{At $|m|=M$, the two continuum roots coincide. The independent solutions are $e^{M\tau}$ and $\tau e^{M\tau}$, so $\Phi(0)=0$ selects $\Phi\propto\tau e^{M\tau}$. The profile of the physical field $\Psi$ derived below is unchanged.} For a constant spinor $\chi$ with $\gamma^\tau\chi=\chi$, the change of variables $\Psi=\cT_M^\dagger\Phi$ in \eqref{eq:change} then gives
\beq
\begin{aligned}
\Phi&\propto \big(e^{|m|\tau}-e^{M\tau}\big)\chi,\\
\Psi&=\cT_M^\dagger\Phi = \i(\partial_\tau-M)\Phi
\propto e^{|m|\tau}\chi.
\end{aligned} \label{eq:continuum-edge-physical-field}
\eeq
The $e^{M\tau}$ term drops out, leaving exactly the profile and chirality of \eqref{eq:edge-mode}, up to an overall normalization. Although $\Phi$ vanishes at $\tau=0$, its normal derivative gives a nonzero $\Psi$ there, with $P_-\Psi=0$ as required by $\mathsf L$.

At the other corners, let $n_\pi\geq1$ be the number of momentum components along $W_0$ equal to $\pi/a$. Using $a^2mM\geq-\beta^2$ and $\beta\leq1$, we have $b-2=4n_\pi+a^2mM\geq4-\beta^2>0$, so \eqref{eq:corner-zero-condition} excludes a boundary zero mode. The mechanism is that the Wilson term gives the species at these corners a positive mass $m_{n_\pi}$, as in \eqref{eq:corner-mass}. Where the mass is negative, the second-order equation has two decaying solutions and the boundary condition selects one. Where the mass is positive, only one solution of each chirality decays, and the boundary conditions forbid both, just as \eqref{eq:norm-identity} forbids a boundary mode for a positive mass.

At a momentum that is not a corner, the part of $\cD_{\rm n}$ acting along $W_0$ and the second term of \eqref{eq:R-lat-boundary} couple $\phi^+$ and $\phi^-$. For $\zeta_\alpha=0$ a simple argument excludes a zero mode at such a momentum. Write $\bar k_i=\sin(ak_i)/a$ and $|\bar k|^2=\sum_i\bar k_i^2$, which is nonzero away from the corners, and let $\Gamma=\sum_i\gamma^i\bar k_i/|\bar k|$. The matrix $\Gamma$ is hermitian, squares to one and anticommutes with $\gamma^\tau$. On a field of momentum $k$ along $W_0$, the difference $\nabla_i$ acts as $\i\bar k_i$. For $\zeta_\alpha=0$ the second term of \eqref{eq:R-lat-boundary} is absent, so that
$
\cK^{\rm lat}_m=\Delta+mM+(M-m)\gamma^\tau\nabla_\tau+\i(M-m)\,|\bar k|\,\Gamma
$.
The field also vanishes on the boundary layer. The difference $\nabla_\tau$ is then antihermitian by \eqref{eq:lat-SBP}, and $\Gamma$ times the sum of the first three terms of $\cK^{\rm lat}_m$ is hermitian. Only the last term therefore contributes to the imaginary part 
$
\Im\,\langle\phi,\Gamma\,\cK^{\rm lat}_m\phi\rangle=(M-m)\,|\bar k|\,\|\phi\|^2
$,
where the inner product is summed over the layers $j\leq-1$. Thus $\cK^{\rm lat}_m\phi=0$ implies $\phi=0$. The same argument applies to $(\cK^{\rm lat}_m)^\dagger$, so neither operator has a zero mode.

\paragraph{The chiral kinetic term.} It remains to discuss whether the zero at $k=0$ describes a propagating chiral fermion. Since $\cK^{\rm lat}_m$ is not hermitian, we need both its right zero mode ($\cK^{\rm lat}_m\phi=0$), which determines the fermion field, and its left zero mode ($\overline\psi\cK^{\rm lat}_m=0$), which determines the conjugate field in the action. Return to $k=0$, so $b=2+a^2mM$. Order the roots of \eqref{eq:roots} as $0\leq y_2<y_1<1$. The right mode has positive chirality and profile
\beq
f_j=\frac{y_1^{-j}-y_2^{-j}}{y_1-y_2}. \label{eq:right-edge-profile}
\eeq
It is normalized here by $f_{-1}=1$. At $\beta=1$ the root $y_2$ is zero and $f_j=y_1^{-j-1}$, as found above.

To find the left mode, we solve $(\cK^{\rm lat}_m)^\dagger\overline\psi{}^\dagger=0$. At $k=0$ the two chirality components decouple. Let $g_j$ denote the negative-chirality profile of $\overline\psi{}^\dagger$ on layer $j$.

Order the rows and columns by $j=-1,-2,\ldots$. In the bulk, changing from $\gamma^\tau=+1$ in \eqref{eq:positive-chirality-recurrence} to $\gamma^\tau=-1$ interchanges the hopping coefficients $1+\beta$ and $1-\beta$. To obtain the first diagonal entry, we must also include the hopping from the first interior layer $j=-1$ ($t=n-1$) to the boundary layer $t=n$. For general $\mu$, the terms in $\Delta$ and $-\i(M-\mu)\cD_{\rm n}$ involving $\phi_n$ give, under \eqref{eq:lat-gauge-collar},
\beq
-\frac{1}{a^2}\phi_n+\frac{M-\mu}{2a}\gamma^\tau\phi_n
=-\frac{\zeta_\alpha}{a^2}\left[1+\frac{a(M-\mu)}{2}\right]P_-\phi_{n-1}, \label{eq:lat-bdy-row}
\eeq
where we used $\phi_n=\zeta_\alpha P_-\phi_{n-1}$ from \eqref{eq:lat-space} and $\gamma^\tau P_-=-P_-$.\footnote{On a finite slab, the corresponding term at $t=1$ has $P_+\phi_1$ in place of $P_-\phi_{n-1}$. For $\zeta_\alpha=0$, the hopping to the boundary layers $t=0,n$ and the second term of \eqref{eq:R-lat-boundary} vanish.}
Setting $\mu=m$ and multiplying by $a^2$ gives the correction $-(1+\beta)\zeta_\alpha$ to the first diagonal entry. At $k=0$, the negative-chirality block is therefore
\beq
\left.a^2\cK^{\rm lat}_m\right|_{\gamma^\tau=-1}=
\begin{pmatrix}
b-(1+\beta)\zeta_\alpha & -(1-\beta) & 0 & 0 & \cdots\\
-(1+\beta) & b & -(1-\beta) & 0 & \cdots\\
0 & -(1+\beta) & b & -(1-\beta) & \ddots\\
0 & 0 & -(1+\beta) & b & \ddots\\
\vdots & \vdots & \ddots & \ddots & \ddots
\end{pmatrix}. \label{eq:left-edge-matrix}
\eeq
Taking the adjoint transposes this real matrix. The equations $a^2 (\cK^{\rm lat}_m)^\dagger g_j=0$ are
\beq
\begin{aligned}
&\big[b-(1+\beta)\zeta_\alpha\big]g_{-1}-(1+\beta)g_{-2}=0 &&(j=-1),\\
&{-}(1-\beta)g_{j+1}+bg_j-(1+\beta)g_{j-1}=0 &&(j\leq-2).
\end{aligned} \label{eq:left-edge-equations}
\eeq
Choose the normalization $g_{-1}=1$. The first equation then fixes the next value,
\beq
g_{-2}=\frac{b}{1+\beta}-\zeta_\alpha. \label{eq:left-edge-initial-values}
\eeq
The equation at $j\leq-2$ in \eqref{eq:left-edge-equations} determines all the remaining values by stepping into the bulk.

We can express this solution using the right profile $f_j$. Both $f_j$ and $f_{j+1}$ satisfy the recurrence in \eqref{eq:left-edge-equations} for $j\leq-2$. The boundary condition and normalization give $f_0=0$ and $f_{-1}=1$, and the right-mode equation at $j=-1$ gives $f_{-2}=b/(1+\beta)$. Thus $f_j-\zeta_\alpha f_{j+1}$ has the required values of $g_{-1}$ and $g_{-2}$. We therefore obtain
\beq
\begin{aligned}
 g_j&=f_j-\zeta_\alpha f_{j+1}
=\frac{(y_1-\zeta_\alpha)y_1^{-j-1}-(y_2-\zeta_\alpha)y_2^{-j-1}}
{y_1-y_2}\quad(j\leq-1).
\end{aligned} \label{eq:left-edge-profile}
\eeq
Since both $f_j$ and $f_{j+1}$ decay into the bulk, this gives one normalizable left mode.

The opposite chirality $\gamma^\tau=+1$ has no decaying adjoint mode: in its bulk recurrence the coefficients $1+\beta$ and $1-\beta$ are exchanged, so its roots are $1/y_1$ and $1/y_2$ (only $1/y_1$ at $\beta=1$), all larger than one.

For small $k$, the terms of $\cK^{\rm lat}_m$ linear in $k$ are $(2\i\beta/a)\gamma^i k_i$ on every row and, from the second term of \eqref{eq:R-lat-boundary}, $(\i \zeta_\alpha/2a)\gamma^\tau\gamma^ik_i$ on the rows of the layer $j=-1$. The Laplacian $\Delta_W$ along $W_0$ changes only at order $k^2$. Expand $\phi$ and $\overline\psi$ in the right and left modes \eqref{eq:right-edge-profile} and \eqref{eq:left-edge-profile}. Using $P_-\gamma^\tau=-P_-$, their low-energy action is the chiral map
\beq
\cK^{\rm lat}_{\rm edge}(k)=\i v\,P_-\gamma^iP_+ k_i+O(|k|^2),\qquad
v \propto 2\beta\sum_{j\leq-1}g_jf_j-\frac{\zeta_\alpha}{2}\,g_{-1}f_{-1}. \label{eq:edge-kinetic}
\eeq
The second contribution to $v$ comes from the spin coupling on the layer $j=-1$. The coefficient $v$ is positive in a region of parameter space.\footnote{With the normalizations $f_{-1}=g_{-1}=1$, summing the geometric series and using $y_1y_2=(1-\beta)/(1+\beta)$ and $y_1+y_2=b/(1+\beta)$ gives
\beq
2\beta\sum_{j\leq-1}g_jf_j-\frac{\zeta_\alpha}{2}
=\frac{(1+\beta)^2(2-\zeta_\alpha b)}{4-b^2}-\frac{\zeta_\alpha}{2}. \label{eq:edge-overlap-sum}
\eeq
Since $1\leq b<2$, the right-hand side is positive precisely when
\beq
\zeta_\alpha<\frac{4(1+\beta)^2}{2b(1+\beta)^2+4-b^2}.
\eeq
It seems that this bound is also enough to exclude a zero mode at a momentum that is not a corner.} This region includes both the continuum limit $a\to0$ at fixed $m$, $M$, and $\alpha$, and the case $\zeta_\alpha=0$.

In correlation functions, the boundary fermion and its conjugate can be represented by $P_+\phi$ and by $\overline\psi P_-$ on the layer $t=n-1$, where the profiles \eqref{eq:right-edge-profile} and \eqref{eq:left-edge-profile} are nonzero.

\subsection{Simple examples}\label{sec:lat-simple-examples}

We illustrate the lattice construction in one dimension, with $\gamma^\tau=1$. The spinor has one component, and the spin coupling $\cR_a$ is absent. The interval example compares the lattice construction with the continuum first-order problem in Section~\ref{sec:first-order}. The circle example compares the phase of the lattice partition function with the APS $\eta$-invariant of the continuum theory.

\paragraph{An interval.} Take $Y=[-\ell,0]$ and free links, as in Section~\ref{sec:first-order}. The continuum second-order operator and its boundary conditions are
\beq
\begin{aligned}
\cK_\mu&=-\partial_\tau^2+(M-\mu)\partial_\tau+\mu M,\\
\Phi(0)&=0,\qquad (\partial_\tau-M)\Phi(-\ell)=0.
\end{aligned}\label{eq:interval-second-order}
\eeq
The minus sign in the Robin condition comes from the outward normal $-\partial_\tau$ at $\tau=-\ell$. On the lattice, \eqref{eq:lat-space} gives $\phi_n=0$ and $\phi_0=\zeta_\alpha\phi_1$, so the independent variables are $\phi_1,\ldots,\phi_{n-1}$.

To display the matrix, write its bulk hopping and diagonal coefficients as
\beq
\beta_\mu=\frac{a(M-\mu)}2,\qquad b_\mu=2+a^2\mu M.\label{eq:simple-example-coefficients}
\eeq
Here $\beta_m=\beta$. In the order $t=1,\ldots,n-1$, the matrix $\cK^{\rm lat}_\mu$ is given by
\beq
a^2\cK^{\rm lat}_\mu=
\begin{pmatrix}
b_\mu-(1+\beta_\mu)\zeta_\alpha & -(1-\beta_\mu) & 0 & \cdots & 0\\
-(1+\beta_\mu) & b_\mu & -(1-\beta_\mu) & \ddots & \vdots\\
0 & -(1+\beta_\mu) & b_\mu & \ddots & 0\\
\vdots & \ddots & \ddots & \ddots & -(1-\beta_\mu)\\
0 & \cdots & 0 & -(1+\beta_\mu) & b_\mu
\end{pmatrix}.\label{eq:interval-lattice-matrix}
\eeq
The first row is the bulk row at $t=1$ with $\phi_0=\zeta_\alpha\phi_1$ substituted, as in \eqref{eq:lat-bdy-row}. The last row is the bulk row at $t=n-1$ with $\phi_n=0$ substituted.

We now evaluate the determinant ratio. Let $d_j(\mu)$ be the determinant of the leading $j\times j$ block of $a^2\cK^{\rm lat}_\mu$, so that $d_{n-1}(\mu)=\det(a^2\cK^{\rm lat}_\mu)$. The first two determinants follow directly from \eqref{eq:interval-lattice-matrix}. For $j\geq3$, expansion along the last row gives the diagonal entry $b_\mu$ times $d_{j-1}(\mu)$, minus the product of the two adjacent off-diagonal entries times $d_{j-2}(\mu)$. Thus
\beq
\begin{aligned}
d_1(\mu)&=b_\mu-(1+\beta_\mu)\zeta_\alpha,\\
d_2(\mu)&=b_\mu d_1(\mu)-(1-\beta_\mu^2),\\
d_j(\mu)&=b_\mu d_{j-1}(\mu)
-(1-\beta_\mu^2)d_{j-2}(\mu),\qquad j\geq3.
\end{aligned}\label{eq:interval-determinant-recurrence}
\eeq
Substituting $d_j(\mu)\propto r^{j-1}$ into the recurrence gives $r^2-b_\mu r+1-\beta_\mu^2=0$. Its two characteristic roots are
\beq
\begin{aligned}
r_{\mu,\pm}
=\frac{b_\mu\pm\sqrt{b_\mu^2-4(1-\beta_\mu^2)}}2
=1+\frac{a^2\mu M}{2}
\pm\frac a2\sqrt{(M+\mu)^2+a^2\mu^2M^2}.
\end{aligned}\label{eq:interval-determinant-roots}
\eeq
The solution is a linear combination of $r_{\mu,+}^{j-1}$ and $r_{\mu,-}^{j-1}$. Matching $d_1(\mu)$ and $d_2(\mu)$ fixes the coefficients and gives the exact finite-lattice determinant
\beq
\begin{aligned}
d_{n-1}(\mu)
={}\frac{r_{\mu,+}-(1+\beta_\mu)\zeta_\alpha}{r_{\mu,+}-r_{\mu,-}}
r_{\mu,+}^{n-1}
-\frac{r_{\mu,-}-(1+\beta_\mu)\zeta_\alpha}{r_{\mu,+}-r_{\mu,-}}
r_{\mu,-}^{n-1}.
\end{aligned}\label{eq:interval-determinant-exact}
\eeq

To take the continuum limit, keep $\ell=na$, $M$, $\mu$ and $\alpha$ fixed. First suppose $M+\mu\ne0$, so the roots separate at order $a$. Equations~\eqref{eq:simple-example-coefficients} and \eqref{eq:interval-determinant-roots}, together with $\zeta_\alpha=1-aM+O(a^2)$, give
\beq
\begin{aligned}
r_{\mu,\pm}&=1\pm\frac a2|M+\mu|+O(a^2),\\
r_{\mu,\pm}-(1+\beta_\mu)\zeta_\alpha
&=\frac a2\big(M+\mu\pm|M+\mu|\big)+O(a^2),\\
r_{\mu,+}-r_{\mu,-}&=a|M+\mu|+O(a^3).
\end{aligned}\label{eq:interval-determinant-expansion}
\eeq
Since $n=\ell/a$, the powers satisfy $r_{\mu,\pm}^{n-1}\to e^{\pm|M+\mu|\ell/2}$. Substituting these limits into \eqref{eq:interval-determinant-exact} yields
\beq
\begin{aligned}
\lim_{\substack{a\to0\\na=\ell}}d_{n-1}(\mu)
=e^{(M+\mu)\ell/2}.
\end{aligned}\label{eq:interval-determinant-limit}
\eeq
For $M+\mu>0$ only the first exponential contributes, while for $M+\mu<0$ only the second contributes.
By continuity in $\mu$, \eqref{eq:interval-determinant-limit} also holds at $\mu=-M$.

Finally, $\det\cK^{\rm lat}_\mu=a^{-2(n-1)}d_{n-1}(\mu)$, and the factors $a^{-2(n-1)}$ cancel in the ratio. Setting $\mu=m$ and $\mu=M$ in \eqref{eq:interval-determinant-limit} gives
\beq
\lim_{\substack{a\to0\\na=\ell}}
\frac{\det\cK^{\rm lat}_m}{\det\cK^{\rm lat}_M}
=e^{-(M-m)\ell/2}.\label{eq:interval-lattice-limit}
\eeq
This is the same result as the one in \eqref{eq:interval-midpoint-partition}, with the midpoint value $R(\tau,\tau)=-\i/2$.

How does the absence of eigenvalues for the first-order operator $\cD_Y=\i\partial_\tau$ with $\Psi(-\ell)=0$ manifest itself on the lattice? Consider the interval matrix \eqref{eq:interval-lattice-matrix} at $\mu=m$, $\beta=1$ and $\zeta_\alpha=0$. Its superdiagonal vanishes, and the boundary value is $\phi_0=0$. Since $b_m=2+a^2mM$, we obtain the backward difference
\beq
\frac{\i a}{2}\big[(\cK^{\rm lat}_m-mM)\phi\big]_t
=\frac{\i}{a}(\phi_t-\phi_{t-1}),\qquad t=1,\ldots,n-1.
\label{eq:interval-first-order-lattice}
\eeq
The matrix $\cK^{\rm lat}_m$ is lower triangular with constant diagonal $b_m/a^2$, so this is its only eigenvalue, of algebraic multiplicity $n-1$. At this eigenvalue the eigenvector equation reduces to $\phi_t=0$ for $t=1,\ldots,n-2$. Only $\phi_{n-1}$ is free, so the eigenspace is one-dimensional and the matrix has no complete eigenbasis for $n>2$. The first-order difference in \eqref{eq:interval-first-order-lattice} has the same eigenspace, and its sole eigenvalue is $\i/a$. Thus no eigenvalue stays finite as $a\to0$ with $na=\ell$ fixed, in agreement with the absence of eigenvalues in the continuum.

\paragraph{A circle with holonomy.} Now take $Y=S^1$ of circumference $L=na$, with sites $t=0,\ldots,n-1$ and periodic spin structure. There are no boundary layers, and every site carries an independent variable. Let the $\U(1)$ holonomy be $\exp(\i\oint_{S^1}A)=e^{\i\theta}$. Since $U_\tau(t)\phi_{t+1}$ transports the field from $t+1$ back to $t$, the links obey
\beq
\prod_{t=0}^{n-1}U_\tau(t)=e^{-\i\theta},\qquad
U_\tau(t)=e^{-\i\theta/n},\label{eq:circle-holonomy}
\eeq
where the second equality is a gauge choice. On the Fourier mode $\phi_t=e^{2\pi\i kt/n}$, the eigenvalue of $\cK^{\rm lat}_\mu$ is
\beq
\kappa_\mu(q_k)=\frac{b_\mu-2\cos q_k+2\i\beta_\mu\sin q_k}{a^2},
\qquad q_k=\frac{2\pi k-\theta}{n},\quad k=0,\ldots,n-1.
\label{eq:circle-lattice-eigenvalues}
\eeq
In particular, $\kappa_M(q_k)=M^2+4\sin^2(q_k/2)/a^2>0$.

The product of these eigenvalues can be evaluated exactly. For $\mu=m,M$, define the two roots of the bulk recurrence by
\beq
y_{\mu,\pm}=\frac{b_\mu\pm\sqrt{b_\mu^2-4(1-\beta_\mu^2)}}{2(1+\beta_\mu)}.
\label{eq:circle-recurrence-roots}
\eeq
They are the roots of \eqref{eq:roots} with the mass replaced by $\mu$ and the momentum along $W_0$ set to zero. The eigenvalue factors as
\beq
\kappa_\mu(q)=-\frac{1+\beta_\mu}{a^2}e^{-\i q}
\big(1-y_{\mu,+}e^{\i q}\big)\big(1-y_{\mu,-}e^{\i q}\big).
\label{eq:circle-eigenvalue-factorization}
\eeq
Using $\prod_{k=0}^{n-1}(1-ye^{\i q_k})=1-y^ne^{-\i\theta}$ and $\beta_M=0$, the common factors cancel in the ratio, giving
\beq
\begin{aligned}
Z^{\rm lat}(S^1,\theta)
=(1+\beta_m)^n 
\frac{\big(1-y_{m,+}^ne^{-\i\theta}\big)\big(1-y_{m,-}^ne^{-\i\theta}\big)}
{\big(1-y_{M,+}^ne^{-\i\theta}\big)\big(1-y_{M,-}^ne^{-\i\theta}\big)}.
\end{aligned}\label{eq:circle-lattice-determinant}
\eeq
This is the exact finite-lattice partition function, including its complex phase.

For the continuum limit, keep $L,m,M$ fixed and take $a\to0$. The two roots and the prefactor satisfy
\beq
\begin{aligned}
\{y_{\mu,+}^n,y_{\mu,-}^n\}&\longrightarrow\{e^{\mu L},e^{-ML}\},\\
(1+\beta_\mu)^n&\longrightarrow e^{(M-\mu)L/2}.
\end{aligned}\label{eq:circle-root-limits}
\eeq
Substituting into \eqref{eq:circle-lattice-determinant} gives
\beq
\begin{aligned}
\lim_{\substack{a\to0\\na=L}}Z^{\rm lat}(S^1,\theta)
=e^{(M-m)L/2}\frac{1-e^{mL-\i\theta}}{1-e^{ML-\i\theta}}
=\frac{\sinh\big((mL-\i\theta)/2\big)}
{\sinh\big((ML-\i\theta)/2\big)}.
\end{aligned}\label{eq:circle-continuum-determinant}
\eeq
For $m<0<M$ and $|m|L,ML\gg1$, its magnitude contains the real bulk vacuum factor $e^{(|m|-M)L/2}$. After removing this factor, the partition function tends to $-e^{\i\theta}$. The exponentially small corrections are of order $e^{-|m|L}$ and $e^{-ML}$.

To compare with the continuum $\eta$-invariant, the same links correspond to $D_\tau=\partial_\tau-\i\theta/L$ acting on periodic functions. Thus
\beq
\cD_Y=\i\partial_\tau+\frac{\theta}{L},\qquad
\lambda_k=-\frac{2\pi k-\theta}{L},\quad k\in\mathbb Z.
\label{eq:circle-continuum-spectrum}
\eeq
For $0<\theta<2\pi$ there is no zero mode. Using the exponential regulator of~\cite{Witten:2019bou} with our normalization of $\eta_D$, the positive and negative eigenvalues give two geometric series,
\beq
\begin{aligned}
\eta_D
&=\frac12\lim_{\epsilon\to0^+}\sum_{k\in\mathbb Z}
\sign(\lambda_k)e^{-\epsilon|\lambda_k|}\\
&=\frac12\lim_{\epsilon\to0^+}
\frac{e^{-\epsilon\theta/L}-e^{-\epsilon(2\pi-\theta)/L}}
{1-e^{-2\pi\epsilon/L}}
=\frac12-\frac{\theta}{2\pi}.
\end{aligned}\label{eq:circle-eta-invariant}
\eeq
Consequently, the lattice construction reproduces the continuum phase,
\beq
\lim_{\substack{m\to-\infty\\M\to+\infty}}
\lim_{\substack{a\to0\\na=L}}
\frac{Z^{\rm lat}(S^1,\theta)}{|Z^{\rm lat}(S^1,\theta)|}
=-e^{\i\theta}=e^{-2\pi\i\eta_D}.
\label{eq:circle-eta-comparison}
\eeq
At $\theta=0$ modulo $2\pi$, the nonzero eigenvalues pair and there is one zero mode. Counting it with $\sign(0)=+1$ gives $\eta_D=1/2$ and the same phase $-1$. For antiperiodic spin structure, replacing $\theta$ by $\theta+\pi$ gives the phase $e^{\i\theta}$.

\section{Discussion}\label{sec:discussion}

\subsection{Summary of results}\label{sec:discussion-summary}

To define chiral fermions in $d$ dimensions, we have constructed the path integral of a massive fermion $\Psi$ on a $(d+1)$-dimensional manifold $Y$ with boundary $W=\partial Y$, using the second-order operators. As in~\cite{Witten:2019bou,Witten:2026ezr}, we impose the local boundary condition $\mathsf L$, given by $P_-\Psi|_W=0$ with $P_\pm=\tfrac12(1 \pm \gamma^\tau)$, where $\gamma^\tau$ is the gamma matrix along the outward normal.
The Lagrangian is $-\overline\Psi(\slashed D_Y+m)\Psi$. Near $W$, the normal coordinate $\tau$ satisfies $\tau=0$ on $W$ and $\tau<0$ inside $Y$. For $m<0$, the localized mode is $\Psi(\tau,x)=\sqrt{2|m|}\,e^{|m|\tau}\chi(x)$, where $\chi$ is a massless chiral fermion on $W$ with $P_-\chi=0$.

The eigenvectors of the Dirac operator $\cD_Y=\i\slashed D_Y$ on $Y$ need not form a complete set and therefore cannot always provide a basis for expanding the fields in the path integral. In the simplest example there are no eigenvectors at all.

Let $\cT_\mu=\cD_Y+\i\mu$, where $\mu$ is an arbitrary mass. The second-order operators are
$
\cK_\mu=\cT_\mu \cT_M^\dagger
$,
where $M > 0$ is a regulator mass. This operator acts on a field $\Phi$ that is related to the original fermion field $\Psi$ by $\Psi=\cT_M^\dagger \Phi$. 
Requiring $\Phi$ to obey $\mathsf L'$ (which is defined by $P_+\Phi|_W=0$) and its image $\Psi$ to obey $\mathsf L$ gives the boundary conditions
\beq
P_+\Phi|_W=0,\qquad (\partial_\tau+M)P_-\Phi|_W=0. \label{eq:summary-boundary-conditions}
\eeq
For $\mu=M$, the positive reference operator $\cK_M=\cT_M\cT_M^\dagger$ supplies a complete orthonormal basis $u_j$ with $\cK_Mu_j=s_j^2u_j$ and $s_j\geq M$. These eigenmodes provide a space of fields shared by the physical fermion with mass $\mu=m$ and a Pauli--Villars regulator with mass $\mu=M$, leading to the determinant ratio
\beq
Z(m;M)=\frac{\det\cK_{m}}{\det\cK_{M}},\qquad (\cK_{\mu})_{jk}=\langle u_j,\cK_\mu u_k\rangle. \label{eq:summary-cutoff-ratio}
\eeq
Since $\det\cK_{M}>0$, this ratio retains the phase of $\det\cK_{m}$. For complete regularization, we considered both multiple Pauli--Villars regulators in the continuum and lattice regularization.

On the lattice, the second-order term naturally supplies Wilson's term. In the bulk, the operator takes the form
\beq
\cK^{\rm lat}_m=(M-m)D_{\rm Wilson}+\cR_a, \label{eq:summary-wilson}
\eeq
where $\cR_a$ supplies the spin coupling to the gauge field. In $D_{\rm Wilson}$, the Wilson term is proportional to the nonnegative lattice Laplacian $\Delta$, with its coefficient fixed by the two masses as $ra/2=1/(M-m)$. In the mass range studied in this paper, the Wilson term gives the doublers positive masses, preventing them from producing unwanted boundary fermions. For $\mu=M$, the regulator operator is $\cK^{\rm lat}_M=\Delta+\cR_a+M^2$ in the bulk.

At the boundary layer $t=n$, where $t$ is the lattice coordinate in the normal direction and takes values $0,\ldots,n$, the physical and regulator lattice fields $\phi$ satisfy
\beq
\begin{aligned}
&P_+\phi_n=0,\\
&\frac{P_-\phi_n-U_\tau(n-1)^\dagger P_-\phi_{n-1}}{a}
=-(1-\alpha)M\,P_-\phi_n-\alpha M\,U_\tau(n-1)^\dagger P_-\phi_{n-1},
\end{aligned} \label{eq:summary-lattice-boundary-conditions}
\eeq
where $\phi_t$ is the field on the layer $t$, $a$ is the lattice spacing, $U_\tau(n-1)$ is the gauge link between the layers $t=n-1$ and $t=n$, and $\alpha$ is a real parameter. These conditions determine the field at $t=n$ from the field at $t=n-1$.
If we take $a\to0$ with the other parameters fixed, we recover the continuum boundary conditions \eqref{eq:summary-boundary-conditions} for any fixed $\alpha$.
Alternatively, we can keep $aM$ of order one as $a\to0$. In particular, choosing $\alpha=1/(aM)$ gives $\phi_n=0$.

The regulator determinant remains positive. Thus the lattice partition function
\beq
Z^{\rm lat}=\frac{\det\cK^{\rm lat}_m}{\det\cK^{\rm lat}_M},\qquad \det\cK^{\rm lat}_M>0, \label{eq:summary-lattice-ratio}
\eeq
retains the phase of the physical determinant, just as in the continuum. 

In the free theory, an isolated boundary supports a single propagating chiral fermion in this mass range. The Pauli--Villars regulator remains gapped, with the square roots of the eigenvalues of $\cK^{\rm lat}_M$ bounded below by $M$.

\subsection{Comparison with other formulations}\label{sec:related-work}

Several ingredients of the present formulation are well known in the literature.

The use of positive second-order operators such as $D^\dagger D$ is familiar in the study of chiral fermions (see, for example,~\cite{Fujikawa:1983bg}). Examples close to the present setting include the analysis of fermions with local boundary conditions~\cite{Sint:1993un} in the formulation of transition amplitudes~\cite{Luscher:1992an}, and the analysis of chiral boundary modes~\cite{Shamir:1993zy,Kaplan:2023pxd}. 
Limitations of the first-order spectrum with $\mathsf L$ were discussed in~\cite{Ivanov:2022jor}. 
Forming second-order determinants with the adjoint of a fixed reference operator has also been discussed in~\cite{Dang:2019byu,Hsu:2026gmi}.

The original domain-wall construction~\cite{Kaplan:1992bt} localizes a chiral fermion where the mass of a higher-dimensional fermion changes sign. The boundary-fermion variant~\cite{Shamir:1993zy} uses an actual end of the lattice and provides a lattice regularization with the local condition $\mathsf L$. More recent work studies chiral fermions on a single connected boundary in the continuum~\cite{Kaplan:2023pxd} and on a finite lattice~\cite{Kaplan:2023pvd} (see also~\cite{Clancy:2024bjb,Aoki:2024bwx,Sen:2026pnm}).
Positive Pauli--Villars denominators are used in both earlier and recent domain-wall formulations~\cite{Furman:1994ky,Kaplan:2024ezz}. Common boundary terms for the physical fermion and its regulator also appear in domain-wall formulations with boundaries in Euclidean time~\cite{Takeda:2010ai,Murakami:2017yzc}.

\paragraph{Comparison with Shamir's boundary-fermion construction.}\label{sec:lat-shamir}

The formulation of~\cite{Shamir:1993zy} is the closest to ours, since it realizes the local boundary condition $\mathsf L$ on a lattice. In both formulations, a massive fermion in the bulk supports a chiral fermion at the boundary. 

Equation \eqref{eq:wilson} exhibits Wilson's operator in our bulk operator, with Wilson parameter $r=2/[a(M-m)]$. In our formulation, choosing $a(M-m)=2$ gives $r=1$, for which the normal-direction terms in $(M-m)D_{\rm Wilson}$ combine into projectors,
\beq
-\frac{2}{a^2}\big[P_-U_\tau(x)\phi(x+a\hat\tau)+P_+U_\tau(x-a\hat\tau)^\dagger\phi(x-a\hat\tau)-\phi(x)\big].
\eeq
In the $r=1$ construction of~\cite{Shamir:1993zy}, the first-order lattice Dirac operator is defined with the same projector structure (see Eq.~(9) of that reference).

In the formulation of~\cite{Shamir:1993zy}, the lattice field $\psi$ occupies only the layers $0<t<n$, so the layers $t=0$ and $t=n$ are missing. With the projector couplings above, omitting the hopping to the missing layers imposes $P_-\psi|_{t=n}=0$ and $P_+\psi|_{t=0}=0$. These are the local conditions $\mathsf L$ at the two ends, with a single chiral mode on each boundary. The boundary condition is therefore supplied by the open end of the Wilson operator itself.

For $r\neq1$, both normal hopping matrices, proportional to $\gamma^\tau-r$ and $\gamma^\tau+r$, are invertible. Omitting the hopping to a missing layer then sets every component of $\psi$ there to zero, rather than imposing a chiral projector. The chiral mode nevertheless persists, as noted in Section~4 of~\cite{Shamir:1993zy} for $r$ close to $1$. For such $r$, the free equation $D_{\rm Wilson}\psi=0$ at zero momentum along $W_0$ has two solutions of the same chirality that decay into the bulk, and the vanishing of $\psi$ on the missing layer selects one combination of them. 

In our formulation, we can interpret the vanishing of both chiral components on the missing layers of~\cite{Shamir:1993zy} for $r\neq1$ in terms of the two boundary conditions on $\phi$. On the layer $t=n$, the projector condition $\mathsf L'$ in \eqref{eq:lat-Lprime} sets $P_+\phi_n=0$. The vanishing of the other component, $P_-\phi_n=0$, is the lattice Robin condition \eqref{eq:lat-robin} at $\alpha aM=1$, for which $\zeta_\alpha=0$ in \eqref{eq:lat-space}. On the layer $t=0$, $\mathsf L'$ sets $P_-\phi_0=0$, and the Robin condition gives $P_+\phi_0=0$. For $a(M-m)<2$, where $r>1$, the condition $P_+\phi_n=0$ selects the combination \eqref{eq:right-edge-profile} of the two decaying solutions of the recurrence \eqref{eq:positive-chirality-recurrence}.

The relation of these two decaying solutions to the physical field is worked out in Section~\ref{sec:lat-chiral}. In the continuum limit $a\to0$ with $m$, $M$, and $\alpha$ held fixed, the two positive-chirality solutions decay into the bulk as $e^{|m|\tau}$ and $e^{M\tau}$, and $\Phi$ is proportional to their difference. The operator $\cT_M^\dagger$ removes the $e^{M\tau}$ contribution, leaving the physical boundary mode $\Psi=\cT_M^\dagger\Phi$ with profile $e^{|m|\tau}$. When $aM$ is of order one, this cancellation is generally incomplete, but the boundary condition still fixes the relative amplitude of the two solutions. Only one chiral boundary mode remains, so we expect the boundary fermion to remain in the same universality class.

The vanishing of the field on a missing layer in~\cite{Shamir:1993zy} cannot be interpreted as a Dirichlet condition on all components in the continuum limit. In the continuum example of Section~\ref{sec:sine-cutoff}, imposing Dirichlet conditions gives a physically nonsensical result. The boundary condition of~\cite{Shamir:1993zy} can be realized in our formulation only when $aM$ is of order one. The freedom in discretizing the Robin condition allows us to choose how the mass term is distributed between the layers $t=n$ and $t=n-1$. Choosing $\alpha$ so that $\alpha aM=1$ gives $\phi_n=0$. With this discretization fixed, maintaining $\phi_n=0$ as $a$ decreases requires $M=1/(\alpha a)$. Taking $a\to0$ with $M$ fixed instead gives the continuum Robin condition $(\partial_\tau+M)P_-\Phi=0$.

\subsection{Numerical evaluation of the \texorpdfstring{$\eta$}{eta}-invariant}\label{sec:numerical-eta}

A natural application of our formulation is the numerical evaluation of the phase $\exp(-2\pi\i\eta_D)$ for fixed background gauge fields. We do not undertake such numerical studies in this paper and leave them to future work.

The proposal of~\cite{Witten:2026ezr} evaluates $\eta_D$ using eigenmodes of a second-order differential operator with the APS boundary condition. A numerical implementation for an anomaly-free two-dimensional $\U(1)$ model was found to agree with exact continuum results~\cite{Misumi:2026ham}.

There are also lattice approaches to the $\eta$-invariant. The phase of the boundary-fermion determinant in an interpolating gauge field was studied as a lattice $\eta$-invariant in~\cite{Tatsumi:1999jlf}, and a related construction of chiral gauge theories was given in~\cite{Kikukawa:2001mw}.

Related studies of anomaly inflow express the APS index through the $\eta$-invariant of a massive domain-wall operator~\cite{Fukaya:2017tsq,Fukaya:2019myi} (see also~\cite{Fukaya:2019qlf,Fukaya:2020tjk,Kobayashi:2021jbn}). Further work evaluates this spectral asymmetry on lattices with curved domain walls~\cite{Aoki:2022aez} and relates lattice Wilson operators to continuum indices using $K$-theory~\cite{Aoki:2024sjc,Aoki:2025gca}.

Lattice calculations also evaluate the exponentiated bulk $\eta$-invariant associated with the Arf--Brown--Kervaire invariant from massive Wilson-fermion Pfaffians~\cite{Araki:2025xly}.

For the interpolated gauge fields of the regularization of~\cite{Hsu:2026gmi,Witten:2026ezr}, the continuum formulation of Section~\ref{sec:sine-cutoff} could be used to evaluate the phase with the local boundary condition $\mathsf L$, instead of the APS boundary condition. On a finite-dimensional space of spinors obeying $\mathsf L'$, the denominator matrix is positive by \eqref{eq:approximation-norm}, so the phase of the determinant ratio comes entirely from the numerator. The approximation criterion \eqref{eq:form-approximation} requires these spaces, as they are enlarged, to approximate each allowed spinor in the norm defined by $q_M$, which controls its derivatives and boundary values. The Robin condition need not be imposed on the basis functions in this application. As for the lattice in Section~\ref{sec:lat-counterterms}, the imaginary part of the logarithm of this ratio requires no counterterm, up to irrelevant operators.

The lattice formulation of Section~\ref{sec:lattice} provides another route through the phase of the determinant ratio~\eqref{eq:Zlat}. It would be interesting to study numerically whether this phase approaches $\exp(-2\pi\i\eta_D)$ in the continuum and large-mass limits.
For such lattice evaluations, we expect admissibility, both on the boundary and in the bulk, to be important for obtaining a good approximation to the continuum phase. This expectation is motivated by the central role of topology, especially bordism, in the continuum analysis. This condition requires the holonomy $U(p)$ around every plaquette $p$ to satisfy $\|1-U(p)\|<\epsilon$ for sufficiently small $\epsilon>0$, and was introduced to define topology on the lattice~\cite{Luscher:1981zq}. Admissible gauge fields also play a central role in the recent constructions of chiral gauge theories in~\cite{Craig:2026soq,Hsu:2026gmi,Witten:2026ezr}.

\section*{Acknowledgements}
The author would like to thank Edward Witten for sharing his paper~\cite{Witten:2026ezr} before submission, as well as his notes.
The work of KY is supported in part by JST FOREST Program (Grant Number JPMJFR2030, Japan), and JSPS KAKENHI (21K03546).
GPT-6, GPT 6.2 Astra, Claude Opus 5.5, and Fable 5.2 have been used throughout this work. The author takes full responsibility for the contents of this paper.

\bibliographystyle{ytphys}
\def\url#1{\href{#1}{#1}}
\bibliography{inspire,other}

\end{document}